\documentclass[aps,prb,twocolumn,showpacs]{revtex4-1}
\usepackage{graphics}
\usepackage{epsfig}
\usepackage{amsmath}
\usepackage{amssymb}
\usepackage{amsfonts}
\usepackage{marvosym }
\usepackage{epstopdf}
\usepackage{pstricks}
\usepackage{bm}
\newcommand{\etal}{\textit{et al.\ }}

\begin{document}
\title{All-electron quasi-particle self-consistent $GW$ band structures
  for SrTiO$_3$ including lattice polarization corrections in different
phases}
\author{Churna Bhandari}
\affiliation{Department of Physics and Astronomy,
  University of Missouri, Columbia Missouri}
\author{Mark van Schilgaarde}
\affiliation{Department of Physics, King's College London, London WC2R 2LS, United Kingdom}
\author{Takao Kotani}
\affiliation{Department of Applied Mathematics and Physics, Tottori University, Tottori 680-8552, Japan}
\author{Walter R. L. Lambrecht}
\affiliation{Department of Physics, Case Western Reserve University,
  Cleveland, OH 44106-7079}
\begin{abstract}
  The electronic band structure of SrTiO$_3$ is investigated in
  the all-electron QS$GW$ approximation. Unlike previous pseudopotential
  based QS$GW$ or single-shot $G_0W_0$ calculations, the gap
  is found to be significantly overestimated compared to experiment.
  After putting in a correction for the underestimate of the screening
  by the random phase approximation in terms of a 0.8$\Sigma$ approach,
  the gap is still overestimated. The 0.8$\Sigma$ approach is discussed and
  justified in terms of various recent literature results including
  electron-hole corrections.
  Adding a lattice polarization correction
  (LPC) in the ${\bf q}\rightarrow0$ limit for the screening of $W$,
  agreement with experiment is recovered. The LPC is alternatively estimated
  using a polaron model. We apply our  approach 
  to the cubic and tetragonal phases as well as a hypothetical
  layered post-perovskite structure and
  find that the LDA (local density approximation)
  to $GW$ gap correction is almost independent of structure. 
\end{abstract}
\maketitle
\section{Introduction} \label{intro}
It is well known that the density functional theory in its
commonly used local density and generalized gradient
approximations (LDA and GGA) does not provide accurate electronic
band structures and in particular underestimates band gaps. This is by now
recognized to be mostly because the Kohn-Sham eigenvalues in this
theory should not be interpreted as one-electron excitations. To calculate
the latter,  a many-body-perturbation theory, including
a dynamical self-energy, such as the $GW$ approximation, provide a much
better justified and more accurate framework. For standard tetrahedral
semiconductors, the $GW$ method has been shown to provide accurate
gaps. Still, this depends on details of the implementation, for example,
all-electron results may differ from pseudopotential results and
the level of self-consistency used in the $GW$ method and its
convergence versus various parameters plays a significant role. For
transition metal and complex oxides, it is still far less clear
how well the $GW$ method performs.  Here we consider  SrTiO$_3$ as
a case study.

We use the all-electron full-potential linearized muffin-tin orbital
(FP-LMTO) implementation\cite{Methfessel,Kotani10}
of the quasiparticle self-consistent (QS) $GW$ method\cite{MvSQSGW,Kotani07}
and compare its results for SrTiO$_3$  with previous results in
literature.\cite{Sponza013,Hamann09,Cappellini00,Takao07}
\section{Literature review}\label{review}
Sponza \etal performed $G_0W_0$ calculations of the band structure
starting from a pseudopotential LDA calculation including
Sr $4s,4p$ and Ti $3s,3p$ semicore states as valence.
They obtain the vertical gap at $\Gamma$ to be
3.76 in good agreement with experiment, whereas their LDA calculation
gave 2.21 eV.
The actual valence band maximum (VBM) at $R$ is slightly higher than at $\Gamma$
resulting in a smaller indirect gap both in LDA and in $GW$.
The focus of their paper is on the
optical dielectric function including electron-hole interaction effects.

Hamann and Vanderbilt (HV) \cite{Hamann09} performed QS$GW$ calculations
using maximally localized Wannier functions (MLWF) to interpolate the
self-energy $\Sigma$ matrix between {\bf k} mesh-points on which the
QS$GW$ is performed. A similar functions is played by the atom centered
muffin-tin-orbitals in our approach. They include only Sr-$4p$ semicore
states as valence electrons. Both these groups used the ABINIT package
but used somewhat different cut-off parameters. Their plane-wave cut-off
for the basis set is similar but HV used a smaller number of unoccupied bands.
They obtained the indirect LDA gap of 1.61 and a GW gap of 3.32 eV. Curiously,
the gap correction of HV (1.71 eV) is larger than that of Sponza \etal (1.55 eV). They did not mention the direct gap at $\Gamma$, but assuming all LDA
calculations considered here get similar value for this difference, we'll use our LDA value (0.44 eV) for the difference between
the VBM at $R$ and $\Gamma$. HV's direct gaps at $\Gamma$ would then amount to
2.05 eV (LDA) and 3.76 eV ($GW$). Thus, these two pseudopotential calculations
are in good agreement with each other in spite of the small changes in
parameter choices.  The main point of HV's paper is that the MLWF interpolation
works well and indicates little change in the Wannier functions extracted
from LDA or GW calculations.

A third pseudopotential based $GW$ calculation by Cappellini \etal \cite{Cappellini00} obtained significantly different results. They also include
Sr $4s,4p$, Ti $3s,3p$ as valence electrons and obtain an LDA gap
at $\Gamma$ of 2.24 eV (indirect $R-\Gamma$ of 1.90)  but  $GW$ gaps
of 5.42 eV ($\Gamma-\Gamma$) and 5.07 eV ($R-\Gamma$). 
The reason for this discrepancy is unclear but presumably is related
to the use of a model dielectric function instead of a consistently
calculated one. Finally, a previous FP-LMTO QS$GW$ calculation by
Kotani \etal,\cite{Takao07,Deguchi16} gives the indirect gap at $\Gamma$ of about 4.25
eV but gave few details.

From the above, it appears from the pseudopotential calculations that the $G_0W_0$ gap is close to that of the QS$GW$ gap, and that both are in good
agreement with experiment. The all-electron QS$GW$ gap however seems to
be about 1 eV larger than experiment.  Here we further investigate this issue.

\section{Methods}\label{methods}
The QS$GW$ approximation as implemented in FP-LMTO was described in detail
in Ref. \onlinecite{Kotani07}. The idea behind the QS$GW$ method is
to make an optimal choice of the $H_0$ Hamiltonian so that its Kohn-Sham
eigenvalues $\epsilon_i$ are as close as possible to the quasiparticle energies
$E_i$. To do this, a hermitian but non-local exchange correlation
potential, specified by its  matrix in the basis of the $H_0$ eigenstates,
\begin{equation}
  \left[V_{xc}^{\Sigma}\right]_{ij}=\frac{1}{2}\mathrm{Re}{\left[\Sigma_{ij}(E_i)+\Sigma_{ij}(E_j)\right]},
\end{equation}
is used in $H_0$. Here, $\Sigma(\omega)$ is the energy
dependent self-energy calculated
from $G_0(\omega)$, the one-electron Green's function
corresponding to $H_0$,  in the single-shot $GW$ approximation: $\Sigma=iG_0W_0$. Starting from an LDA $H_0$, $\Sigma$ is calculated, $V_{xc}^{\Sigma}-V_{xc}^{LDA}$ is added to $H_0$, a new $G_0$ calculated and so on till self-consistency.
The reasons behind this approach and differences from fully self-consistent
 sc$GW$ are discussed in Refs. \onlinecite{Kotani07,Takao07,Ismail-Beigi17}.

\begin{figure}
  \includegraphics[width=9cm]{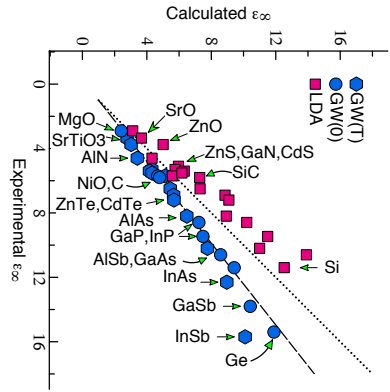}
  \caption{QSGW (blue) and LDA (red) optical dielectric
    constant $\epsilon_\infty$
compared to experiment.  Experimental data for GW(T) is taken at room temperature;
for GW(0) 0K data is used.  LDA data for NiO is not shown because
it is off the scale; similarly the narrow gap semiconductors InAs,
GaSb, InSb and Ge where the LDA gap is negative.
The dotted line would indicate perfect agreeement
between calculation and experiment, while the dashed line corresponds
to 20 \% underestimate by the calculation.\label{figeps}}
\end{figure}

\begin{table*}
  \caption{LDA, QS$GW$ and QS$G\tilde{W}$ gaps, gap corrections and their ratios from Ref. \onlinecite{PasquarelloChen15}, 
    sc$GW$+vertex from Ref. \onlinecite{Kutepov17}\label{tabwp}. B and D refer to two different vertex correction
  schemes, see text.}
  \begin{ruledtabular}
    \begin{tabular}{lcccccc|cccccccc}
      & \multicolumn{3}{c}{$E_g$} & \multicolumn{2}{c}{$\Delta E_g$} &Ratio& \multicolumn{3}{c}{$E_g$}& \multicolumn{3}{c}{$\Delta E_g$} &  \multicolumn{2}{c}{Ratio}\\
      & LDA & QS$GW$ & QS$G\tilde{W}$ & QS$GW$-LDA & QS$G\tilde{W}$-LDA& $\frac{\Delta E_g^{\mathrm{QS}G\tilde W}}{\Delta E_g^{\mathrm{QS}GW}}$ & sc$GW$ & B  & D & sc$GW$-LDA & B-LDA & D-LDA & B & D \\ \hline
MgO &	4.65&	9.29&	8.30&	4.64&	3.65& 	0.79 & 9.31 & 8.24 & 7.96 & 4.66 & 3.59 & 3.31 & 0.77 & 0.71 \\
NiO&	1.05&	4.97&	4.40&	3.92&	3.35&	0.85 & \\
TiO$_2$	&1.90&	4.22&	3.73&	2.32&	1.83&	0.79 &\\
Cu$_2$O& 0.53&	2.65&	2.12&	2.12&	1.59&	0.75 &\\
ZnO&	0.85&	4.61&	3.42&	3.76&	2.57&	0.68 &\\
C&	4.22&	6.4&	5.9&	2.18&	1.68&	0.77 & 6.15 & 5.8 & 5.73 & 1.93 &1.58 & 1.51 & 0.82 & 0.78  \\
SiC&	1.39&	2.9&	2.52&	1.51&	1.13&	0.77 & 2.89 & 2.52& 2.42 & 1.50 & 1.13& 1.03 & 0.75 & 0.69 \\
GaAs&	0.43&	1.75&	1.51&	1.32&	1.08&	0.75 & 2.27 & 1.80 & 1.72 & 1.84 &1.37 & 1.29 & 0.74 & 0.70 \\
BN&	4.53&	7.51&	6.67&	2.98&	2.14&	0.74 & \\
LiCl&	6.52&	10.98&	9.87&	4.46&	3.35&	0.74 & \\
Si& 	0.57&	1.47&	1.3&	0.9&	0.73&	0.76 & 1.55 & 1.32 & 1.26 & 0.98 & 0.75 & 0.69 & 0.77 & 0.70\\
LiF\footnote{From SMK\cite{Shishkin07}}&    9.28&  15.90 & 14.50 &  6.62 & 5.22 & 0.79   & 16.3 & 15.02 & 14.39 & 7.02 & 5.74 & 5.11 & 0.82& 0.73 \\
Ge&	0.00&	0.96&	0.82&	0.96&	0.82&	0.76 &\\
AlP&	1.60&	3.1&	2.77&	1.5&	1.17&	0.75 & 2.84 & 2.53 & 2.44 & 1.24 & 0.93 & 0.84 & 0.75 & 0.68 \\
CdS&	1.21&	3.41&	2.74&	2.2&	1.53&	0.75 & \\ \hline
Average&    &       &       &      &        &  0.76 &&&&&&&0.78&0.72\\
Stdv&       &       &       &      &        &  0.04 &&&&&&&0.04&0.04 \\ 
    \end{tabular}
  \end{ruledtabular}
\end{table*}

For tetrahedral semiconductors, this approach provides systematically  a
$\sim$20 \% overestimate of the gap due to the underestimate of the
dielectric screening in the random phase approximation (RPA) which
does not include electron-hole effects and thus misses ladder diagrams
in the evaluation of the irreducible polarization propagator
$\Pi^0=-iG_0\times G_0$, which determines $W$ through
$W=(1-v\Pi^0)^{-1}v$, where $v$ is the bare Coulomb interaction and a
simplified symbolic operator notation is used. This has led to
the adoption of a universal 0.8$\Sigma$ correction factor.\cite{Chantis06,Chantis08,Deguchi16}
This is illustrated in Fig. \ref{figeps} which shows the typical underestimate
of screening by QSGW to be 20 \% as indicated by the dashed line.
Although it is not clear {\sl a priori} that this also applies to
oxides we adopt a similar correction factor here.

It is interesting that $\epsilon_\infty$ predicted by the LDA is in
sometimes better agreement with experiment.  This can be attributed to a
fortutitous cancellation of errors: missing ladder diagrams tend to
cause $\epsilon_\infty$ to be underestimated, while the LDA's gap
underestimate contributes an error of the opposite sign.  There is no
universal pattern, however, as is already apparent in the data shown in
Fig. \ref{figeps}. Where gap errors are severe the LDA severely overestimates the $\epsilon_\infty$.  For example in NiO, the $\epsilon_\infty^\mathrm{LDA}{>}30$.

Further justification for the 0.8$\Sigma$ correction factor can be obtained
from the work of Shishkin, Marsman and Kresse (SMK)\cite{Shishkin07} and Wei and Pasquarello (WP), \cite{PasquarelloChen15} who added an exchange-correlation
kernel to the screening of the polarization function
$\tilde\Pi=[1-(v+f_{xc})\Pi^0]^{-1}$ using the nanoquanta kernel or a bootstrap kernel respectively. We will refer to their approach as QS$G\tilde W$. 
Although these kernels primarily address the
${\bf q}\rightarrow0$ and static ($\omega=0$)
behavior and might thus not capture the full
extent of the electron-hole effects on renormalizing the screening in $W$,
and have received some critical discussions,\cite{Rigamonti15} it
is useful in the present context to analyze  how much they  affect the gaps
for a variety of materials. 
Analyzing the data in Table I in WP, part of which is reproduced here
in Table \ref{tabwp} with additional analysis, we find that
$[E_g(\mathrm{QS}G\tilde W)-E_g(\mathrm{LDA})]/[E_g(\mathrm{QS}GW)-E_g(\mathrm{LDA})]$ has an average value of about 0.76 with standard deviation of 0.04 with the largest deviation for NiO, where it is 0.85 and ZnO, 0.68. ZnO, is a notably difficult material to converge
and SMK's values for the QS$GW$ and QS$G\tilde{W}$ gaps would give 0.77.
NiO is  a well-known strongly correlated material and a deviation here is not too unexpected. We note that multiplying the self-energy operator $\Sigma$
by 0.8 is not exactly the same as correcting the gap shift by 0.8.
A slightly larger gap reduction typically occurs.

Very recently, Kutepov\cite{Kutepov16,Kutepov17} introduced a
way to solve Hedin's full set of equations\cite{Hedin65,Hedin69}
beyond the $GW$ approximation using systematic diagrammatic approximations
for the vertex function. 
First of all, his results show fully self-consistent sc$GW$ results differ only slightly from the QS$GW$ results and tend to overestimate the band gaps by
a similar amount. 
Secondly, he used two different self-consistency schemes which both
introduce vertex corrections both in $G$ and $\Pi$. The results of his
scheme B, which in his notation only includes a first correction to the vertex
$\Gamma_1$,  are close to those of SMK and WP where comparison for the
same material is possible (Si, LiF, GaAs, SiC, BN, MgO) 
while his most advanced scheme  including the full $\Gamma_{GW}$ vertex,
give a somewhat larger reduction of the gap. These are also shown in Table \ref{tabwp}.
Viewed as percentage of the
sc$GW$-GGA (or LDA) correction they give correction factors of about 0.78 and
0.72 respectively when averaged over various cases.  As an example
for MgO, his sc$GW$ gap is 9.31 and his schemes B and D
give 8.24, 7.96 eV while WP 's QS$GW$ and QS$G\tilde W$ give 9.29, 8.30 eV
and SMK obtain 9.16 eV, 8.12 eV respectively.  The scheme D
agrees almost perfectly with experiment when a lattice-polarization
correction of 0.15 eV added to the experimental volume but the latter
may be somewhat underestimated.\cite{Lambrecht17}
In any case, these results also support that the electron-hole correction
effects beyond RPA amount to about a 20 \% reduction of the QS$GW$ gap
correction beyond LDA or GGA. 

Besides the electron-hole corrections discussed until now,
we also consider a lattice-polarization correction as suggested
by Botti and Marques (BM)\cite{Botti13} and revisited recently in Ref. \onlinecite{Lambrecht17}. The idea here is that for
strongly ionic materials, with large LO-TO phonon splittings,
the $W$ in the long-wave length limit $W({\bf q}\rightarrow0,\omega)$ should include the effects of the ionic displacements on the macroscopic
dielectric constant. The macroscopic dielectric constant
enters the calculation of $\Sigma$ in the special treatment of
the ${\bf q}\rightarrow0$ region in the convolution integral over {\bf k}-space:
\begin{eqnarray}
\Sigma^c_{nm}({\bf k},\omega)&=&\frac{i}{2\pi}\int d\omega'\sum_{\bf q}^{BZ}\sum_{n'}^{all} 
G_{n'n'}({\bf k}-{\bf q},\omega-\omega')\nonumber \\
&&\sum_{\mu\nu}
W^c_{\mu\nu}({\bf q},\omega')e^{-i\delta\omega'} \nonumber \\
&&\langle\psi_{{\bf k}n}|\psi_{{\bf k}-{\bf q}n'}E_\mu^{\bf q}\rangle
\langle E_\nu^{\bf q}\psi_{{\bf k}-{\bf q}n'}|\psi_{{\bf k}m}\rangle \label{eqsig}
\end{eqnarray}
Here, a two-particle mixed product interstitial-plane-wave basis set
$E_\nu$ diagonalizing the bare Coulomb interaction matrix is used\cite{Kotani07,Friedrich10} and $W^c$
is the correlation part of $W$, subtracting the bare exchange.
The need for a special treatment of the ${\bf q}\rightarrow0$ region
arises from the integrable divergence of the
Coulomb interaction ($\propto1/q^2$)
and is here treated using the modified offset-$\Gamma$ method,\cite{Kotanijpsj}
which in turn is closely related to the analytic 
${\bf k}\cdot{\bf p}$ scheme of Friedrich \etal\cite{Friedrich10}
This involves the macroscopic
dielectric tensor ${\bf L}(\omega)$, in their notation
${\bf e}_{\bf k}^T{\bf L}(\omega){\bf e}_{\bf k}$. The projection along
unit vectors ${\bf e}_{\bf k}$ takes care of the non-analytic (orientation dependent) nature of the ${\bf k}\rightarrow0$ limit and fully takes into
account any possible anisotropies depending on the crystal structure.
It is this macroscopic dielectric tensor, usually written
$\boldsymbol{\varepsilon}(\omega)$ which needs to be modified to take into
account the lattice polarization effect.  This is most easily done by
means of a Lyddane-Sachs-Teller factor:
\begin{equation}
\frac{\varepsilon_{tot}^{\alpha}({\bf q}\rightarrow 0,\omega)}{\varepsilon_{el}^{\alpha}({\bf q}\rightarrow 0,\omega)}=\prod_m\frac{\left(\omega^\alpha_{LOm}\right)^2-\omega^2}{\left(\omega_{TOm}\right)^2-(\omega+i0^+)^2}. \label{eqLST} 
\end{equation}
where the superscript $\alpha$ denotes a projection direction
of the tensor,
($\varepsilon^\alpha={\bf e}_{\alpha}^T\boldsymbol{\varepsilon}{\bf e}_{\alpha}$).
It is clear from this expression that the correction goes
to zero for $\omega\gg\omega_L$. In practice we only include it
for $\omega=0$ to avoid the necessity for a careful integration
mesh right near the phonon frequency poles.
As discussed in
Ref. \onlinecite{Lambrecht17} the BM  approach gives the
long-range or Fr\"ohlich contribution to the Fan-part of the zero-point motion
electron-phonon correction of the gap. 
The ${\bf q}$-point integration mesh that needs to be used is a subtle
issue discussed in Lambrecht \etal\cite{Lambrecht17}. The
strength  of this contribution, applied only at ${\bf q}=0$ for convenience,
can be estimated from the polaron length scale,
$a_P=\sqrt{\hbar/2m_* \omega_L}$
  with $m_*$ the band-edge effective mass and $\omega_L$ the relevant LO-phonon
  frequency. We will discuss later how to apply this in the present case with multiple phonons
  and a degenerate VBM not occurring at $\Gamma$. The polaronic point of view allows
  us to make an independent estimate of the corresponding gap reduction. 

\section{Computational Details}\label{comp}
We employ a generalized FP-LMTO method\cite{Methfessel,Kotani10}
as implemented in the Questaal package.\cite{questaal}
The basis set is specified  by two sets of parameters, the smoothing radii $R_{sm}$ and decay lengths ($\kappa$) of smoothed Hankel function envelope functions.\cite{Bott98} For SrTiO$_3$  we include ($spdf$, $spd$) for Sr, ($spd$, $spd$) for Ti and ($spd$,$sp$) for O atoms respectively. These indicate the angular momenta included for each $\kappa$. The envelope functions are augmented inside the spheres in terms of solutions of the Schr\"odinger equation and their energy derivative up to an augmentation cut-off of $l_{max}=4$. In addition, calculations are made with and without the $4p$ ($3p$) local orbitals inside the spheres for Sr and (Ti).

The Brillouin zone integration {\bf k}-point convergence and other convergence
parameters of the method were carefully tested  for cubic SrTiO$_3$ and
similar criteria were adopted for the tetragonal and orthorhombic phases. We also tested result with a larger {\bf k}-point mesh and found the band gap is converged within 0.05 eV. Specifically, we used a $4\times4\times4$ un-shifted mesh for the Brillouin zone sampling, along with the tetrahedron method for the cubic cases in the LDA self-consistent charge convergence and  for the calculation of the $\Sigma$ in $GW$. For the tetragonal phase, the  unit cell is larger along  the $c$-direction
than in-plane by a factor $\sqrt{2}$.  Thus, we use accordingly smaller number of {\bf k}-points, $4\times4\times3$ for both LDA and QS$GW$ calculations.

For the self-consistency cycle, the charge density and the total energy are converged within the tolerance of $10^{-5}$ $e/a_0^3$ and $10^{-5}$ Ry respectively. For QS$GW$, after several convergence test calculations, we settled
the cut-off above which the self-energy matrix is approximated by an
average diagonal value, $\Sigma_{cut}$= 3 Ry, including self-energy
calculations up to 3.5 Ryd, the interstitial 
plane wave cut-off energy for basis functions $E_{cut}(\psi_{G})$=2.6 Ry
and for the auxiliary basis $E_{cut}(\psi_{coul})$=2.8 Ry
respectively. In QS$GW$, the self-consistent iteration was carried until the change in $\Sigma$ was less than $10^{-4}$ Ry.
\section{Crystal structures}\label{structure}
\begin{figure*}
a)\includegraphics[scale=0.4]{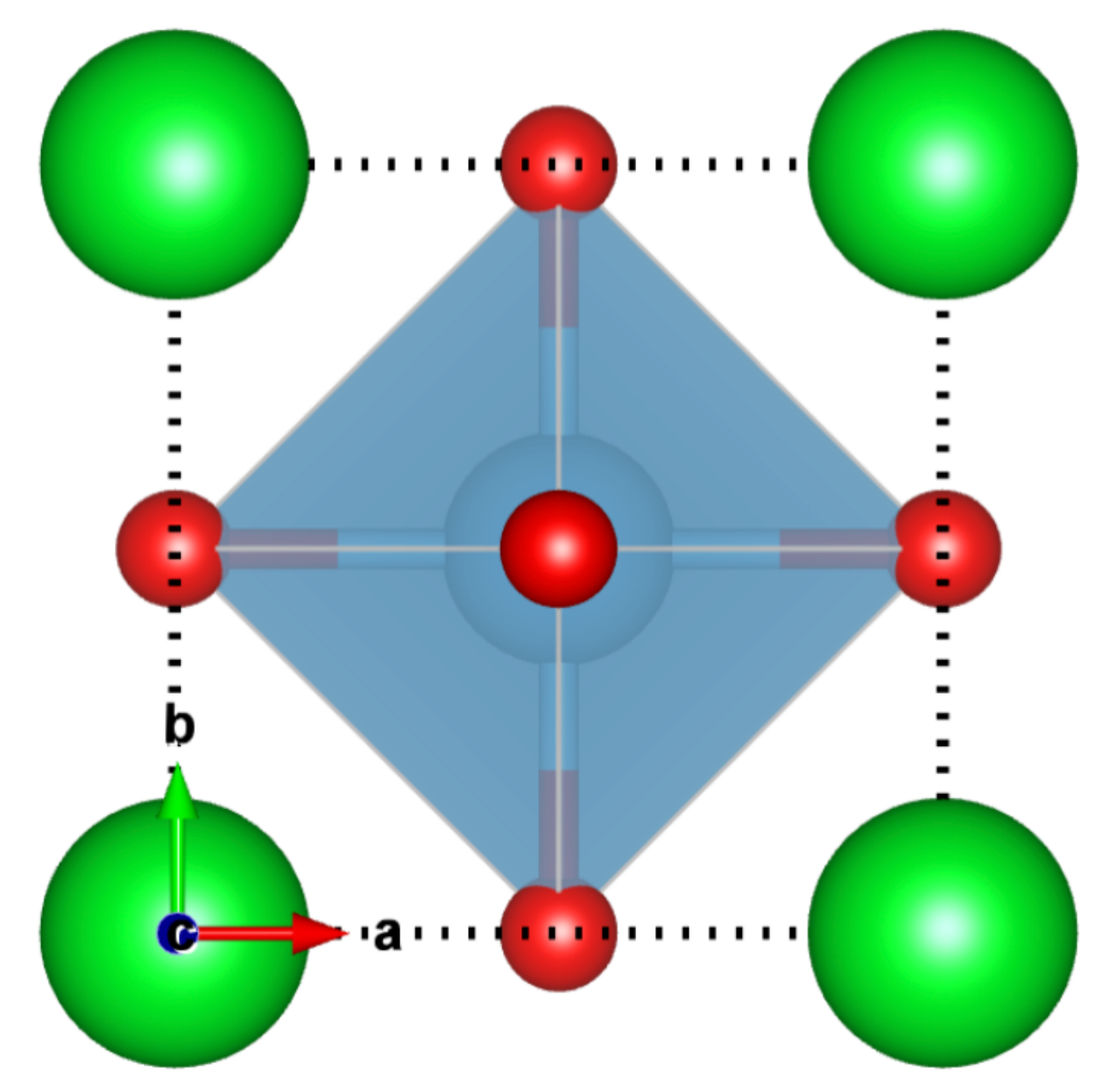}b)\includegraphics[scale=0.4]{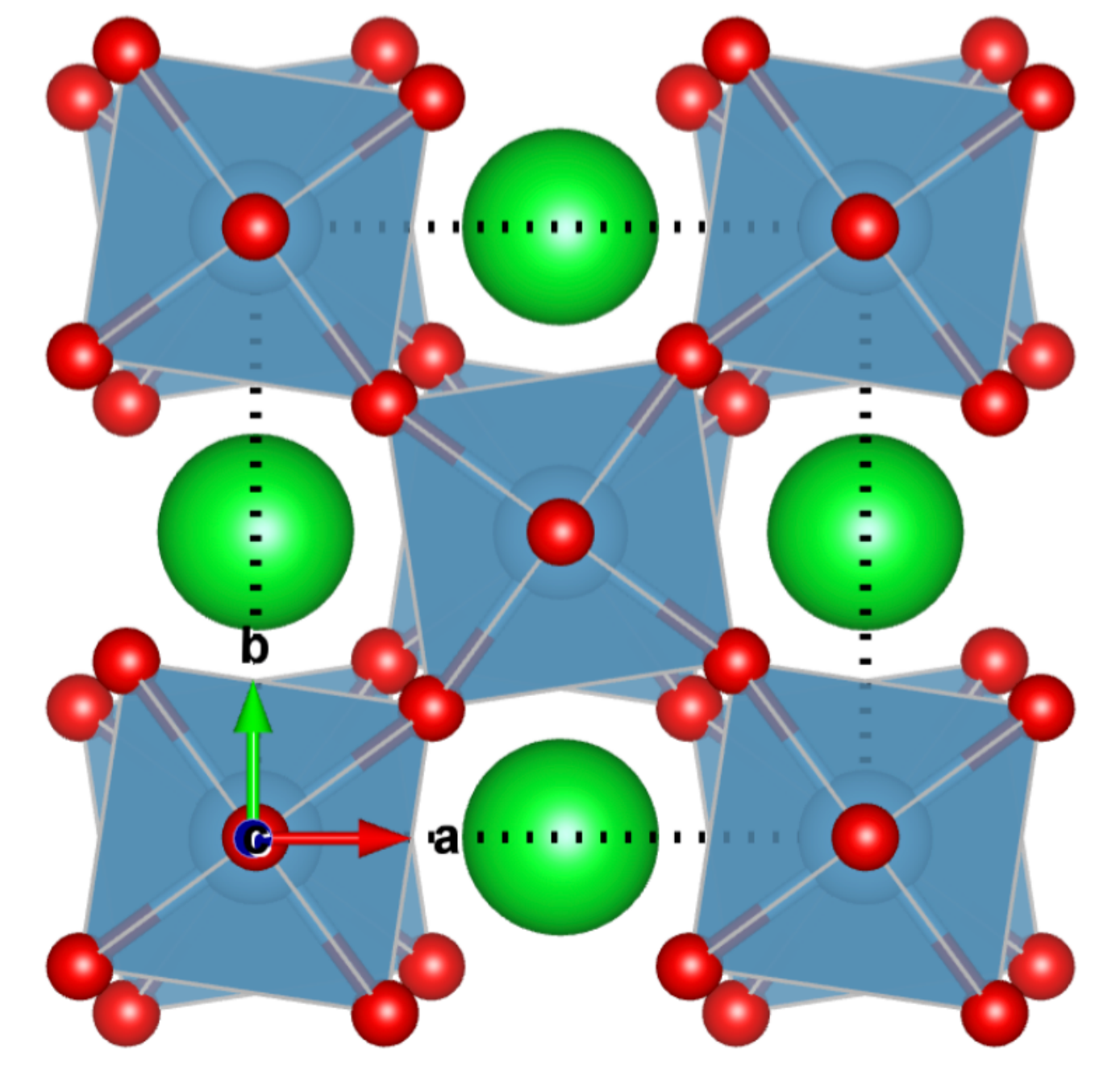}\hspace*{0.5cmc)}\includegraphics[scale=0.4]{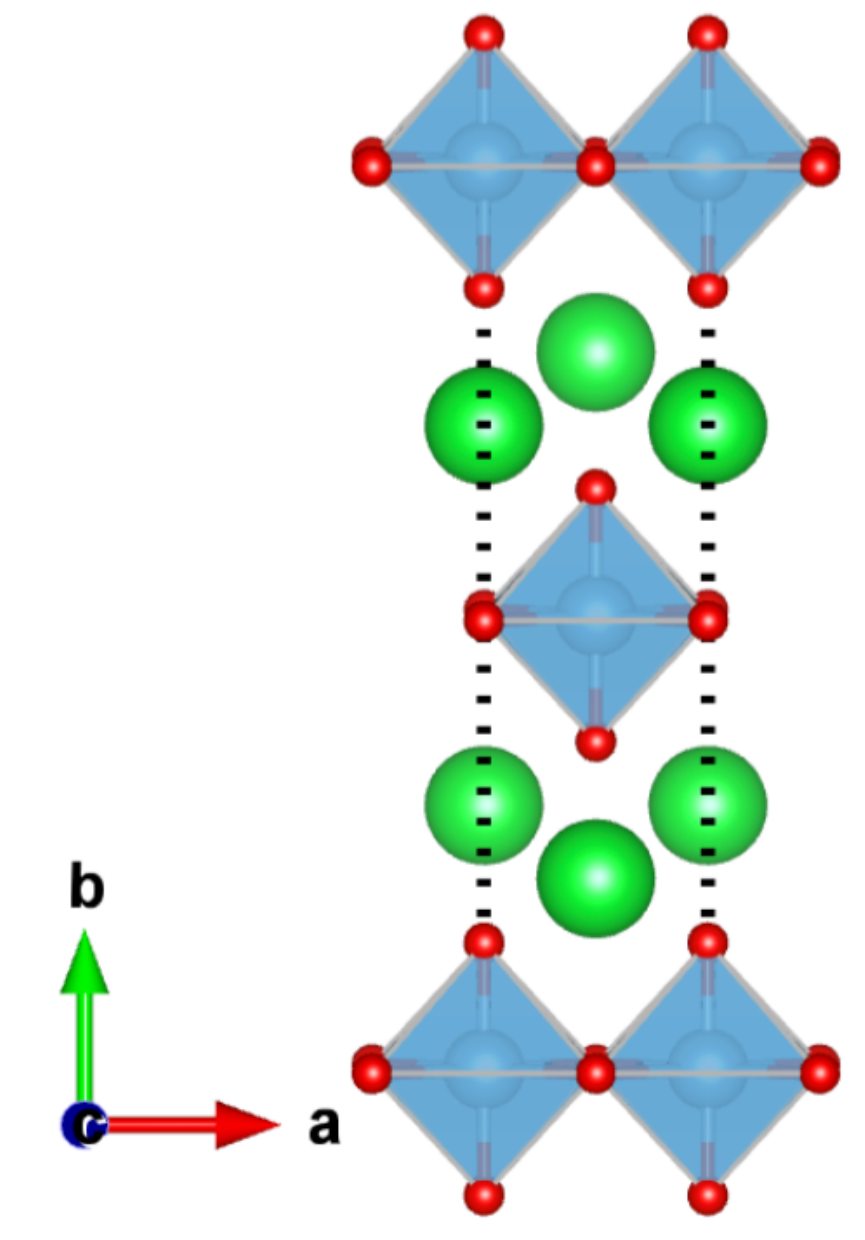}
\caption{Crystal structures of SrTiO$_3$ a) cubic,  b) tetragonal $I4/mcm$, and c) layered orthorhombic CaIrO$_3$ view from the z-axis.}\label{figstruc}
\end{figure*}

We consider the cubic and tetragonal anti-ferro-electrically
distorted (AFD) $I4/mcm$ structure occurring at low temperature.
In addition we consider the layered orthorhombic CaIrO$_3$ structure, suggested
to occur at high pressures by Cabaret \etal\cite{Cabaret07} and also
known as the post-perovskite structure. 
Although we will show elsewhere\cite{Bhandaristostruc} that this structure is unlikely
to occur because it has a higher equilibrium lattice volume and
much higher total energy, it is of interest to see how the
$GW$ gap corrections compare in such different structures. 
Fig. \ref{figstruc} shows the crystal structures for cubic, tetragonal and orthorhombic from left to right respectively. Table \ref{tab.1} summaries the
structural parameters used in the calculations, such as the lattice constants
and Wyckoff positions. The relaxed lattice constant for the cubic phase in LDA is 3.86 \AA~  which is only 1 \% underestimated relative to the experiment.
\begin{table}
\begin{ruledtabular}
  \caption{Experimental lattice constants (except for orthorhombic structure) in \AA ~and Wyckoff position of atoms SrTiO$_3$ in various  structures for cubic and tetragonal
 ($c/a=1.414$ and $w$=0.241.}\label{tab.1}
\begin{tabular}{lccl}
Symmetry & Lattice constant & Wyckoff position \\
\hline
&  & Sr(0,0,0)$\Rightarrow$ 1a  \\
 Cubic &a=3.905\footnote{Expt. A. Yu. Abramov and {\sl et al.}\cite{Abramov95}}&Ti(0.5,0.5,0.5)$\Rightarrow$ 1b\\ 
  & &O(0.5,0.5,0.0)$\Rightarrow$ 3c\\ 
& &Sr(0,0.5,0.25)$\Rightarrow$ 4b \\
Tetragonal  &a=5.507\footnote{Expt. by W. Jauch and {\sl et al.}\cite{Jauch99}}, c=7.796 & Ti(0,0,0)$\Rightarrow$ 4c&\\
AFD &  &O(0,0,0.25)$\Rightarrow$ 4a & \\
 & &(w,0.5+w,0)$\Rightarrow$ 8h & \\
Orthorhombic &a=3.01  &Sr(0,0.2906,0.25)$\Rightarrow$ 4c\\
post-perovskite &b= 12.77 &Ti(0,0,0)$\Rightarrow$ 4a\\
CaIrO$_3$ &c= 5.87&O(0,0.4939,0.25)$\Rightarrow$ 4c\\  
 & &O(0,0.1391,0.999)$\Rightarrow$ 8f\\               
\end{tabular}
\end{ruledtabular}
\end{table}
\section{Results}\label{results}
\subsection{Cubic STO}
\begin{figure}[h]
\includegraphics[width=8cm]{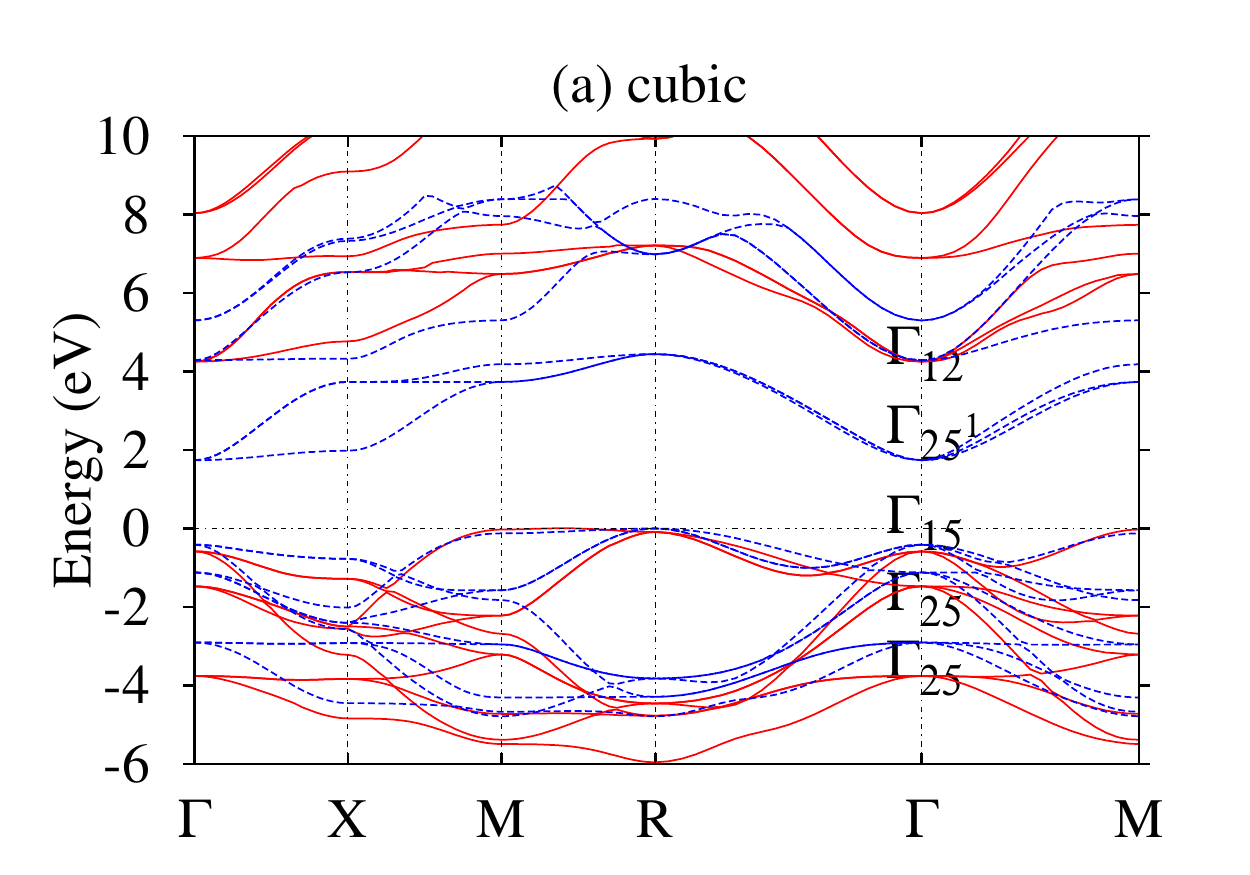}
\includegraphics[width=8cm]{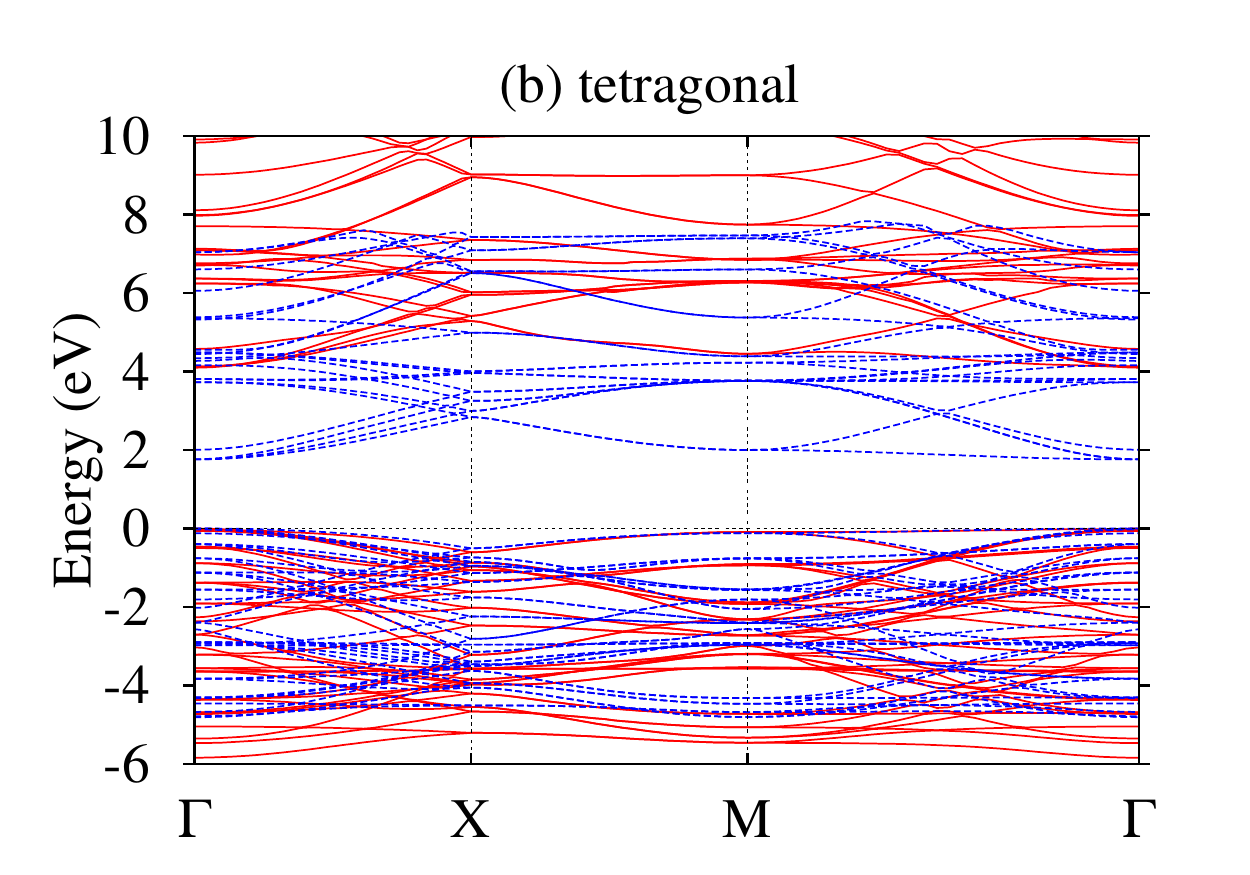}
 \includegraphics[width=8cm]{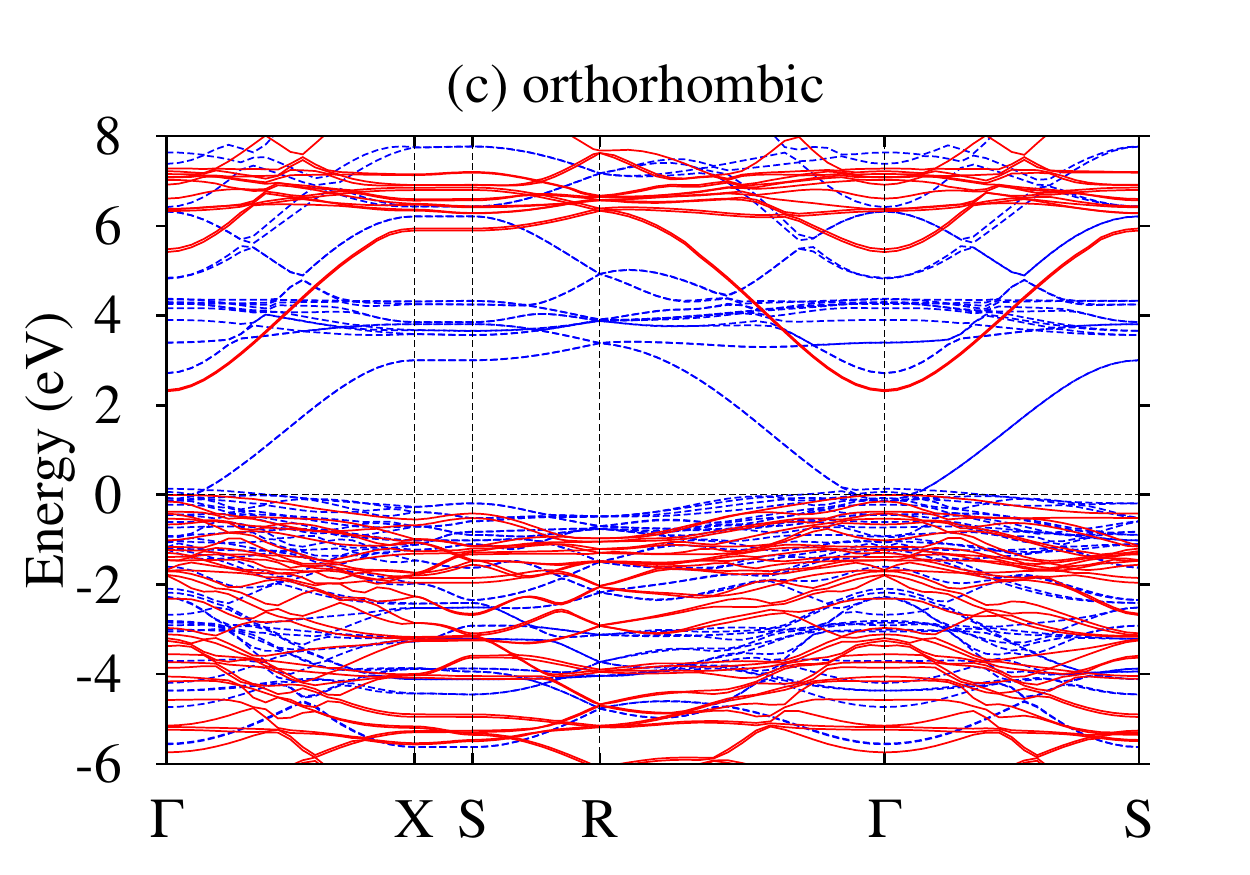}
\caption{(Color-online) Band structure of a) cubic and b)tetragonal SrTiO$_3$ c) layered orthorhombic, LDA: blue dashed and QS$GW$: red solid lines. \label{figgwlda}}
\label{figbnd}
\end{figure}

In Fig. \ref{figbnd}a we show the band structure of
cubic SrTiO$_3$ in the full QS$GW$ approach compared with LDA.
A few states at $\Gamma$ are symmetry labeled for later reference. 
In Table \ref{tabgapscubic} we summarize the gaps and various other
band structure parameters in different approximations.
In Table \ref{tabstates} we show how the different approximations
affect other band states relative to the VBM. This allows us to
assess to what extent  the $GW$ correction can be approximated by
a ${\bf k}$ and state independent scissor shift.

\begin{table*}
\begin{ruledtabular}
  \caption{Band gap in eV of cubic SrTiO$_3$. Here no semi-core
    means Sr$-4p$ and Ti$-3p$ semi-core states are not included.}\label{tabgapscubic}
\begin{tabular}{lclllccccc}
basis-sets&${\bf k}$ point& LDA & $G_0W_0$ &QS$GW$ &$0.8\Sigma$-QS$GW$&QS$GW$+LP & QS$GW$+LP$+0.8\Sigma$& Expt.\\ \hline

with Sr$_{4p}$,Ti$_{3p}$&$\Gamma-\Gamma$ &2.15,2.24\footnote{First principle calculation by Cai Meng-Qui \etal\cite{Meng04,Gupta04}},2.21$^b$&4.1 &4.83,3.76\footnote{Sponza \etal\cite{Sponza013}}5.42\footnote{Cappellini \etal\cite{Cappellini00} }&4.14&4.28 & 3.7 &3.75$^d$\\

no semi-core & &2.08  &3.85 & 4.31 &3.8& 3.83 &3.5& \\

with Sr$_{4p}$,Ti$_{3p}$&$R-\Gamma$ &1.74 &3.65 & 4.34,5.07$^{c}$,3.32\footnote{Hamann and Vanderbilt\cite{Hamann09}} &3.64&3.78&3.2& 3.25\footnote{Fundamental gap in the valence electron-loss spectroscopy \cite{van01,Bauerle78}},3.2\footnote{Fundamental absorption edge in reflectivity measurement of SrTiO$_3$\cite{Cardona65}}\\
no semi-core\footnote{Without local orbital VBM is at $R$} & & 1.65 &3.37 &3.82&3.3& 3.35 &3 &\\\\
with Sr$_{4p}$,Ti$_{3p}$& $\Gamma$-VBM&1.74 & 3.52 & 4.25 &3.52 & 3.69 & 3.15 &\\

\end{tabular}
\end{ruledtabular}
\end{table*}

First, we see that our LDA gap agrees quite well with other LDA (or GGA) calculations. 
Second we see that the $G_0W_0$ gap is significantly lower than the QS$GW$ gap. Third, unlike
the pseudopotential calculations reviewed in Sec. \ref{review}, the QS$GW$ gap significantly
overestimates the gap. Even if we use the $0.8\Sigma$ approach, they are still larger than experiment.
It is only when we add both the $0.8\Sigma$ and lattice polarization correction, that we recover
the experimental values.  We also note that the $0.8\Sigma$ approach
actually reduces the QSGW-LDA indirect (direct) gap shifts by about a factor 0.73 (0.74).  
In agreement with other calculations and already correctly described in
LDA, the indirect $R-\Gamma$ gap is about 0.4 eV lower than the lowest direct $\Gamma-\Gamma$ gap. The VBM between $R-M$ is very flat and in QS$GW$ the
actual VBM lies actually in between R and M and is 0.09 eV above that at R.
Finally we see that the semi-core levels play a more  important role in QS$GW$
than in LDA. Neglecting them, the gap would be only 0.07 eV lower in LDA but is 0.5 eV lower in
QS$GW$ or still 0.2 eV lower in the final LST and $0.8\Sigma$ corrected case. 

\begin{table*}
  \caption{Polaron band shift estimates. The estimate in each row corresponds to a given LO-TO
    phonon pair. The final row gives the sum of them. The masses are in units of free electron mass, the phonon frequencies in cm$^{-1}$, the polaron lengths in Bohr, the band shifts in meV.\label{tabpolaron}}
  \begin{ruledtabular}
    \begin{tabular}{cccccccccccccc}
      $m_{le}$ & $m_{he}$ & $m_e$ & $m_{hh}$ & $m_{lh}$ & $m_h$ & $\omega_L$ & $\omega_T$ & $a_{Pe}$& $a_{Ph}$ & $\epsilon_\infty$ & $\Delta E_c$ & $\Delta E_v$ & $\Delta E_g$ \\ \hline
     0.33 &2.65 & 1.10& 5.08 & 0.67 & 2.14 & 795 & 547 & 11.19& 8.03& 5.52 & 58 & 81 & 134 \\
     0.33&2.65&1.10 & 5.08 & 0.67 & 2.14 & 474 & 170 & 14.49& 10.40&5.52& 74 & 103 & 177 \\
     0.33 &2.65&1.10 & 5.08 & 0.67 & 2.14  & 171 & 91  & 24.12 & 17.32& 5.52&37 & 51  & 88 \\ \hline
      &&       &      &      &      &     &     &      &&   Total& 169 & 235 & 404 \\
    \end{tabular}
  \end{ruledtabular}
\end{table*}

\subsection{Polaron estimates}
Next, we discuss the lattice polarization correction to the gap in detail. 
The zero-point motion correction contains a contribution from the long-range Fr\"ohlich type of
electron phonon coupling. The latter is arguably the largest electron-phonon coupling correction
for a strongly ionic material with large LO-TO splitting because the other electron-phonon coupling
effects tend to be smaller than 0.1 eV except for systems with all light atoms.
To estimate it we follow the approach of Nery and Allen.\cite{NeryAllen16} The main point is that
the Fr\"ohlich electron-phonon coupling behaves as $1/q$ and hence near band edges where the
band difference $E_n({\bf k}+{\bf q})-E_n({\bf k})$, which enters the denominator in
the Allen-Heine-Cardona expression for the electron-phonon self-energy,
gives  a divergent contribution. Nery and Allen showed how it can be integrated analytically
when a simple  effective mass approximation is used for the bands.
The length scale for the polaron effect
is $a_P=\sqrt{\hbar/2m_*\omega_L}$ and if we assume we need to integrate the singular behavior only over a region in ${\bf q}$-space of size $1/a_P$ as upper limit, then the polaron shift of a band is
given by\cite{Lambrecht17}
\begin{eqnarray}
\Delta E_n({\bf k})&=&-\alpha_P\hbar\omega_L/2 \nonumber \\
&=& -\frac{e^2}{4 a_P} \left(\frac{1}{\varepsilon_\infty}-\frac{1}{\varepsilon_0}\right),\nonumber \\
&=& -\frac{e^2}{4 a_P\varepsilon_\infty}\left(1-\frac{\omega_T^2}{\omega_L^2}\right).
\end{eqnarray}
In other words it essentially the change in the Coulomb interaction calculated at the polaron length scale due to the change in screening from only electronic screening to electron  plus lattice screening. The extra factor 2 arises from the choice of cut-off in {\bf q}-space and we have
written the change in macroscopic inverse dielectric constants using the Lyddane-Sachs-Teller
relation. In this way, for a given LO-TO phonon pair, we have a separate contribution
from each phonon, since both $a_P$ and the dielectric constant factor depend on the phonon
considered. We can thus estimate the effect for each phonon and add  them, thereby generalizing
Nery and Allen's simple model to the case of multiple phonons.  In SrTiO$_3$, there are
three optically active phonons.

The second point is that this predicts a correction near each band edge. The conduction
band at $\Gamma$ and the VBM at $R$ are both three-fold degenerate and anisotropic, so to
apply the theory in its simple form, we need to average the effective masses  in some way
to extract the polaron length scale. At both points we could exploit the cubic
symmetry to write a Kohn-Luttinger type of effective Hamiltonian. In our previous work\cite{Lambrecht17}
for simple di-atomic cubic compounds, we just used an average of the heavy and light masses in the
cubic direction, according to the corresponding band's degeneracy. Following the same
approach here, the band structure shows that it would be appropriate to use
$m_h=(m_{hh}+2m_{lh})/3$ for holes and the same for the electrons, $m_e=(m_{he}+2m_{le})/3$
where we use the masses in the $\Gamma-X$ and $R-M$ directions, which are both simple cubic $x$ directions.
Thus,we obtain separate electron and hole polaron length scales $a_{Pe}$
and $a_{Ph}$. Since the latter only provide estimates of the {\bf q}-space integration region, it
is not too crucial how we perform the average, although we recognize this is at present a
limitation of the approach. The VBM at $R$ can be seen to be rather flat and in fact in $GW$
the maximum moves away from $R$ toward $M$. We use the masses extracted from our QS$GW$ bands
without the lattice polarization correction. 
The hole polaron length scale is significantly shorter than for the electrons, predicting
stronger polaronic effects for holes. This agrees with the finding in other work of
self-trapped hole-polarons.\cite{Janotti14,Erhart14}

\begin{figure}
  \includegraphics[width=8cm]{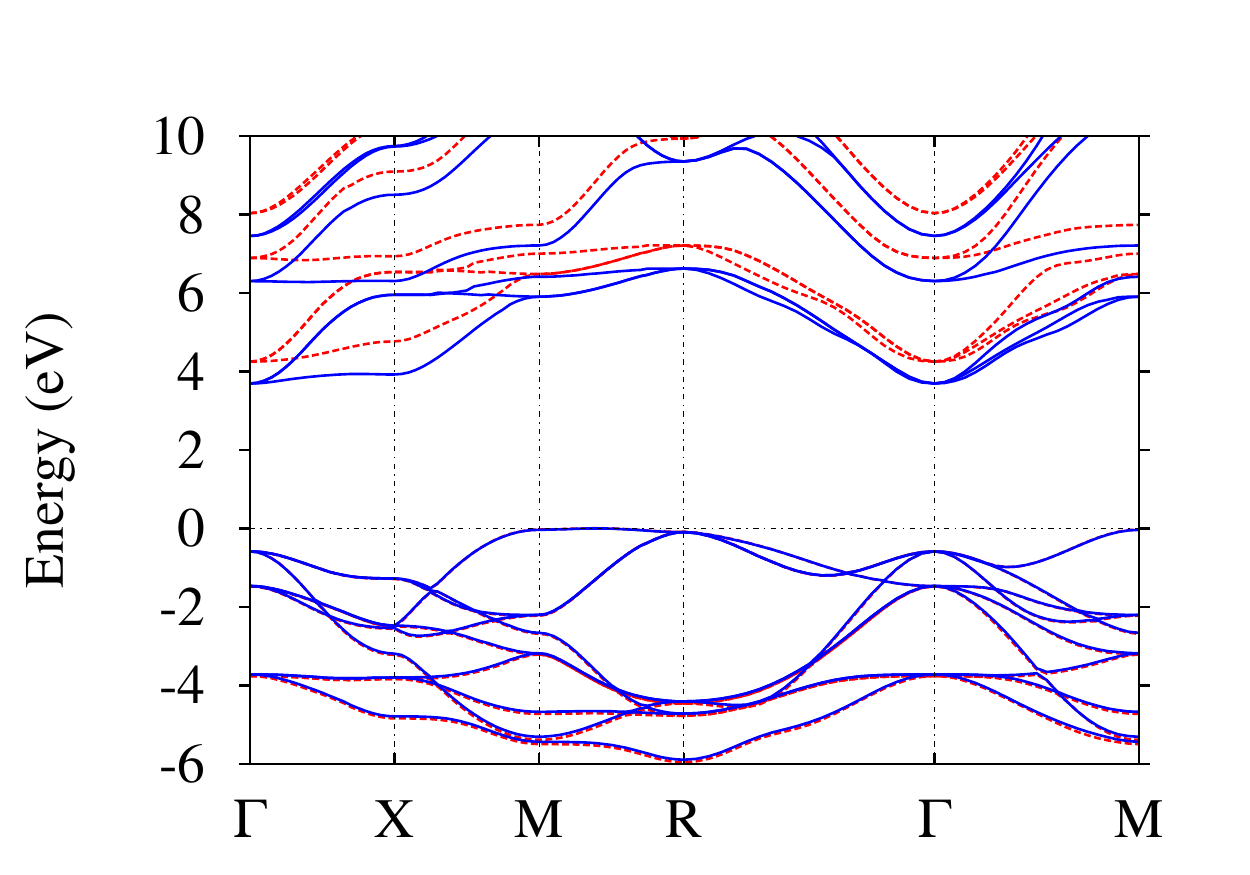}
  \caption{Comparison of the band structure of SrTiO$_3$ in QS$GW$ (red dashed) and
    in QS$GW$ with lattice polarization correction  using BM-approach with
    $4\times4\times4$ mesh (blue solid).
    \label{figbndlp}}
\end{figure}

The results of this approach and the corresponding parameters are summarized in Table \ref{tabpolaron}.
We can see that the conduction band is predicted to shift less than the valence band as expected and the total gap correction
is predicted to be 404 meV, which we really should round off to 0.4 eV. The shortest polaron
length scale corresponds to holes for the largest phonon frequency and is 8 Bohr. This corresponds to a q-space region
of about 1/6 of the Brillouin zone.  Our estimate using the BM approach in Table
\ref{tabgapscubic} used a $4\times4\times4$ mesh and gives
a contribution to the zero-point motion or lattice polarization correction of $-0.55$ eV. This is already rather
close to the polaron estimate. 
With a $6\times6\times6$ mesh we obtain $-0.25$ eV. These bracket the polaron estimate of Table \ref{tabpolaron}. We can thus conservatively
conclude that the lattice polarization correction amounts to $0.3\pm0.1$ eV in good agreement between the Nery-Allen like
estimate (0.4) and the BM approach.   When we add this to the $0.8\Sigma$
result we obtain a gap of 3.24 eV for the indirect gap in excellent agreement with experiment. We note that if we apply the lattice polarization correction
using the BM approach with a $4\times4\times4$ mesh but then apply the $0.8\Sigma$ correction, the LPC shift is also reduced by 0.8, and 
becomes $0.4$ eV.  We can see that in this approach the correction is almost a constant shift and hence the
indirect gap correction is the same as the direct gap correction. Because of the approximate nature of these estimates,
we have not separately evaluated the polaron approach to the VBM at $\Gamma$ which would give the direct gap.
In principle, the polaronic effect also should enhance the band mass by a factor $(1+\alpha_P/6)$ but it is not clear that the
BM-method captures this more subtle effect. In fact, we find the bands to shift almost rigidly as can
be seen in Fig. \ref{figbndlp}.

\subsection{Other band structure features}
Turning to other band features than the gap, summarized in Table \ref{tabstates}
we see that Sr-$4p$ states lie significantly closer to
the VBM than the Ti $3p$ semicore states and hence play a more important role. We can see that
the shifts of these states are also sensitive to the $0.8\Sigma$ and LPC corrections
and amount to  about 2 eV for Sr-$4p$
and 4 eV for Ti $3p$. As expected, the farther away from the VBM, the larger is the quasiparticle
self-energy shift.  In the conduction band we see that the higher lying $\Gamma_{12}$ state has almost the
same shift from LDA (about 1.4 eV) as the $\Gamma_{25'}$ CBM. In the valence band the shifts
are smaller and progressively larger as we go deeper in the VBM.

\begin{table*}
\begin{ruledtabular}
  \caption{Various band differences: specific states (symmetry labeling as in
    Fig. \ref{figgwlda}a) in the conduction band relative to the
    CBM, and in the upper valence band relative to the VBM.  The position of the
    semicore states Sr$_{4p}$, Ti$_{3p}$, are also  with respect to the VBM at $R$.\label{tabstates}}
\begin{tabular}{l|ll|lcc|ccc}
 &$\Gamma_{25^{'}}$-CBM &$\Gamma_{12}$-CBM &$\Gamma_{15}$-VBM &$\Gamma_{25}$-VBM &$\Gamma_{25}$-VBM &Sr$_{4p}$ states &Ti$_{3p}$ states\\
 \hline
 LDA &1.74 &4.29 &-0.41 &-1.12 &-2.91 &-14.66 &-32.55\\
 $G_0W_0$&3.53 &6.29 &-0.58 &-1.37 &-3.49 &-17.19 &-36.48\\
 QS$GW$ &4.34 &6.98 &-0.49 &-1.39 &-3.67 &-17.51 &-37.38\\
 $0.8\Sigma$-QS$GW$ &3.54 &6.17 &-0.49&-1.32 &-3.64 &-16.9 &-36.74\\\
 QS$GW$+LP &3.78 &6.4 &-0.49 &-1.37 &-3.62 &-17.51 &-37\\
 QS$GW$+LP$+0.8\Sigma$&3.2&5.71 &-0.49 &-1.31 &-3.5 &-16.9 &-36.4\\
\end{tabular}
\end{ruledtabular}
\end{table*}

\subsection{Tetragonal structure}
The band structure for the tetragonal structure is shown in Fig. \ref{figbnd}b.
In the tetragonal material we see a similar large shift of the band gap by $GW$.
To understand this band structure, we note that the tetragonal unit cell
is rotated by 45$^\circ$ and has $a_t=\sqrt{2}a_c$ as in-plane lattice constant.
Thus the Brillouin zone (BZ) of the cubic structure is folded into a smaller BZ
with the $\Gamma-M$ of the tetragonal BZ corresponding to half the $\Gamma-X$
of the cubic BZ. The high symmetry points correspond to 
$M=(1/2,1/2,0)$ and $X=(1/2,0,0)$ with respect to their respective reciprocal
lattice vectors. Similarly the $\Gamma-X$ of the tetragonal BZ is half the $\Gamma-M$ of the cubic BZ. One can clearly see the folding in half of the bands with additional small gaps opening due to the breaking of the symmetry by the slight
rotation of the octahedra. We can see that VBM which in the cubic case 
and in QSGW occurs between $M-R$ ($R=(1/2,1/2,1/2)$), where the band dispersion
is very flat, is folded on to the tetragonal BZ $\Gamma$ point
and the gap becomes direct. 
\begin{table}
\begin{ruledtabular}
\caption{Band gap of tetragonal SrTiO$_3$ in eV at $\Gamma-\Gamma$.}\label{tab.3}
\begin{tabular}{clllll}
  basis-sets & LDA & QS$GW$ &QS$GW$&QS$GW$&Ref. \\
             &     &        & +LPC &$0.8\Sigma$+LPC & \\ \hline
 with Sr$_{4p}$,Ti$_{3p}$&1.76 &4.1 &3.88&3.27& 4.03\footnote{$^{,b}$ By Heifets and {\sl et al.} \cite{Heifets06} using a hybrid DFT-Hartree-Fock (HF) approach }\\
no semi-core & 1.72 & 3.7&3.47&3.04& \\
\end{tabular}
\end{ruledtabular}
\end{table}

\subsection{Hypothetical layered orthorhombic structure}
Although, the CaIrO$_3$ structure, proposed\cite{Cabaret07}  for SrTiO$_3$ as a potential high-pressure structure, can be shown to be unstable,\cite{Bhandaristostruc}
it is of interest to see how the GW gap correction changes with such a large change in structure. This structure has edge-sharing octahedra in  layers separated by Sr,
rather than corner sharing octahedra. In the LDA, the band gap becomes zero as can be seen in Fig. \ref{figbnd}c. The very different band dispersion in this case
results from the direct Ti-d to Ti-d interactions between much closer Ti atoms in the layer. 
In the QS$GW$ method the gap becomes 2.32 eV which is not too different from the gap correction 2.68 eV in cubic perovskite. The gap correction is found to be almost the same as in the cubic or tetragonal structures.  Similar screening reduction or $0.8\Sigma$ corrections
and lattice polarization corrections should apply here but are not further pursued at this point. 

\section{Conclusions}
In this paper we reviewed the status of the QS$GW$ method for a prototypical complex transition metal oxide like SrTiO$_3$ in the perovskite structure.
We found that all-electron QS$GW$ results obtained by means of the FP-LMTO implementation give a significant overestimate of the gap compared to experiment
in contrast to PAW or pseudopotential based $GW$ approaches. This indicates a compensation of errors in the latter. We base this on the observation
that for  a large family of materials, the under-screening of $W$ in the RPA amounts to about 20 \%  and can hence be accommodated by using the 0.8$\Sigma$ approach.
This evidence is based both on the
comparison of dielectric constants in QS$GW$ with experiment and on recent
calculations\cite{Shishkin07,PasquarelloChen15}
which go beyond the RPA by including an exchange correlation kernel 
in the calculation of $W$ or adding vertex corrections directly\cite{Kutepov17} and it is found to apply to both tetrahedrally bonded semiconductors and  various oxides and ionic compounds.  The second important
correction to the gap is the lattice polarization correction. This is part of the zero-point motion correction due to electron-phonon coupling
and more specifically is its dominant contribution in strongly ionic materials arising from the long-range Fr\"ohlich part of the electron-phonon coupling.
Two independent estimates of this effect were made: one based on the polaron theory and one on the Botti-Marques approach of multiplying the
macroscopic dielectric constant at ${\bf q}=0$ by a Lyddane-Sachs-Teller factor along with a suitable ${\bf q}$-mesh sampling based itself on the polaron length scale
which determines the strength of the effect. The two estimates are found to be in good agreement with each other.
We find that both the  electron-hole interaction effects which reduce $\Sigma$ by about 20 \% and the lattice-polarization corrections are required to obtain good
agreement with experimental gaps in cubic perovskite SrTiO$_3$. As for the structural dependence of the QS$GW$ corrections, we find that the gap correction in
tetragonal STO is very close to that in cubic STO and the bands are essentially folded according to the rotation of the octahedra, which leads to a
doubling of the cell and rotation of the BZ by 45$^\circ$. This happens to fold the $R$ point of the BZ onto the $\Gamma$-point an hence the indirect lowest gap
becomes then direct. Due to the similarity in band states, we expect it to be pseudo-direct in the sense that no strongly optical transitions will correspond to this
direct gap. Even for a very different hypothetical structure with edge-sharing octahedra, we find very similar gap corrections by QS$GW$, which shows that the gap
corrections are rather insensitive to structure. 

\acknowledgments{This work was supported by the US Department of Energy, Office of Science, Basic
Energy Sciences under grant No.  DE-SC0008933.  Calculations made use of the High Performance
Computing Resource in the Core Facility for Advanced Research Computing at Case Western Reserve
University.  MvS was supported by EPSRC CCP9 Flagship Project No. EP/M011631/1.}

\bibliography{srto,gw,lmto,dft,lst}

\begin{thebibliography}{38}%
\makeatletter
\providecommand \@ifxundefined [1]{%
 \@ifx{#1\undefined}
}%
\providecommand \@ifnum [1]{%
 \ifnum #1\expandafter \@firstoftwo
 \else \expandafter \@secondoftwo
 \fi
}%
\providecommand \@ifx [1]{%
 \ifx #1\expandafter \@firstoftwo
 \else \expandafter \@secondoftwo
 \fi
}%
\providecommand \natexlab [1]{#1}%
\providecommand \enquote  [1]{``#1''}%
\providecommand \bibnamefont  [1]{#1}%
\providecommand \bibfnamefont [1]{#1}%
\providecommand \citenamefont [1]{#1}%
\providecommand \href@noop [0]{\@secondoftwo}%
\providecommand \href [0]{\begingroup \@sanitize@url \@href}%
\providecommand \@href[1]{\@@startlink{#1}\@@href}%
\providecommand \@@href[1]{\endgroup#1\@@endlink}%
\providecommand \@sanitize@url [0]{\catcode `\\12\catcode `\$12\catcode
  `\&12\catcode `\#12\catcode `\^12\catcode `\_12\catcode `\%12\relax}%
\providecommand \@@startlink[1]{}%
\providecommand \@@endlink[0]{}%
\providecommand \url  [0]{\begingroup\@sanitize@url \@url }%
\providecommand \@url [1]{\endgroup\@href {#1}{\urlprefix }}%
\providecommand \urlprefix  [0]{URL }%
\providecommand \Eprint [0]{\href }%
\providecommand \doibase [0]{http://dx.doi.org/}%
\providecommand \selectlanguage [0]{\@gobble}%
\providecommand \bibinfo  [0]{\@secondoftwo}%
\providecommand \bibfield  [0]{\@secondoftwo}%
\providecommand \translation [1]{[#1]}%
\providecommand \BibitemOpen [0]{}%
\providecommand \bibitemStop [0]{}%
\providecommand \bibitemNoStop [0]{.\EOS\space}%
\providecommand \EOS [0]{\spacefactor3000\relax}%
\providecommand \BibitemShut  [1]{\csname bibitem#1\endcsname}%
\let\auto@bib@innerbib\@empty
\bibitem [{\citenamefont {Methfessel}\ \emph {et~al.}(2000)\citenamefont
  {Methfessel}, \citenamefont {van Schilfgaarde},\ and\ \citenamefont
  {Casali}}]{Methfessel}%
  \BibitemOpen
  \bibfield  {author} {\bibinfo {author} {\bibfnamefont {M.}~\bibnamefont
  {Methfessel}}, \bibinfo {author} {\bibfnamefont {M.}~\bibnamefont {van
  Schilfgaarde}}, \ and\ \bibinfo {author} {\bibfnamefont {R.~A.}\ \bibnamefont
  {Casali}},\ }in\ \href@noop {} {\emph {\bibinfo {booktitle} {Electronic
  Structure and Physical Properties of Solids. The Use of the LMTO Method}}},\
  \bibinfo {series} {Lecture Notes in Physics}, Vol.\ \bibinfo {volume} {535},\
  \bibinfo {editor} {edited by\ \bibinfo {editor} {\bibfnamefont
  {H.}~\bibnamefont {Dreyss{\'e}}}}\ (\bibinfo  {publisher} {Berlin Springer
  Verlag},\ \bibinfo {year} {2000})\ p.\ \bibinfo {pages} {114}\BibitemShut
  {NoStop}%
\bibitem [{\citenamefont {Kotani}\ and\ \citenamefont {van
  Schilfgaarde}(2010)}]{Kotani10}%
  \BibitemOpen
  \bibfield  {author} {\bibinfo {author} {\bibfnamefont {T.}~\bibnamefont
  {Kotani}}\ and\ \bibinfo {author} {\bibfnamefont {M.}~\bibnamefont {van
  Schilfgaarde}},\ }\href {\doibase 10.1103/PhysRevB.81.125117} {\bibfield
  {journal} {\bibinfo  {journal} {Phys. Rev. B}\ }\textbf {\bibinfo {volume}
  {81}},\ \bibinfo {pages} {125117} (\bibinfo {year} {2010})}\BibitemShut
  {NoStop}%
\bibitem [{\citenamefont {van Schilfgaarde}\ \emph {et~al.}(2006)\citenamefont
  {van Schilfgaarde}, \citenamefont {Kotani},\ and\ \citenamefont
  {Faleev}}]{MvSQSGW}%
  \BibitemOpen
  \bibfield  {author} {\bibinfo {author} {\bibfnamefont {M.}~\bibnamefont {van
  Schilfgaarde}}, \bibinfo {author} {\bibfnamefont {T.}~\bibnamefont {Kotani}},
  \ and\ \bibinfo {author} {\bibfnamefont {S.}~\bibnamefont {Faleev}},\ }\href
  {\doibase 10.1103/PhysRevLett.96.226402} {\bibfield  {journal} {\bibinfo
  {journal} {Phys. Rev. Lett.}\ }\textbf {\bibinfo {volume} {96}},\ \bibinfo
  {pages} {226402} (\bibinfo {year} {2006})}\BibitemShut {NoStop}%
\bibitem [{\citenamefont {Kotani}\ \emph
  {et~al.}(2007{\natexlab{a}})\citenamefont {Kotani}, \citenamefont {van
  Schilfgaarde},\ and\ \citenamefont {Faleev}}]{Kotani07}%
  \BibitemOpen
  \bibfield  {author} {\bibinfo {author} {\bibfnamefont {T.}~\bibnamefont
  {Kotani}}, \bibinfo {author} {\bibfnamefont {M.}~\bibnamefont {van
  Schilfgaarde}}, \ and\ \bibinfo {author} {\bibfnamefont {S.~V.}\ \bibnamefont
  {Faleev}},\ }\href {\doibase 10.1103/PhysRevB.76.165106} {\bibfield
  {journal} {\bibinfo  {journal} {Phys.Rev. B}\ }\textbf {\bibinfo {volume}
  {76}},\ \bibinfo {eid} {165106} (\bibinfo {year}
  {2007}{\natexlab{a}})}\BibitemShut {NoStop}%
\bibitem [{\citenamefont {Sponza}\ \emph {et~al.}(2013)\citenamefont {Sponza},
  \citenamefont {V\'eniard}, \citenamefont {Sottile}, \citenamefont
  {Giorgetti},\ and\ \citenamefont {Reining}}]{Sponza013}%
  \BibitemOpen
  \bibfield  {author} {\bibinfo {author} {\bibfnamefont {L.}~\bibnamefont
  {Sponza}}, \bibinfo {author} {\bibfnamefont {V.}~\bibnamefont {V\'eniard}},
  \bibinfo {author} {\bibfnamefont {F.}~\bibnamefont {Sottile}}, \bibinfo
  {author} {\bibfnamefont {C.}~\bibnamefont {Giorgetti}}, \ and\ \bibinfo
  {author} {\bibfnamefont {L.}~\bibnamefont {Reining}},\ }\href {\doibase
  10.1103/PhysRevB.87.235102} {\bibfield  {journal} {\bibinfo  {journal} {Phys.
  Rev. B}\ }\textbf {\bibinfo {volume} {87}},\ \bibinfo {pages} {235102}
  (\bibinfo {year} {2013})}\BibitemShut {NoStop}%
\bibitem [{\citenamefont {Hamann}\ and\ \citenamefont
  {Vanderbilt}(2009)}]{Hamann09}%
  \BibitemOpen
  \bibfield  {author} {\bibinfo {author} {\bibfnamefont {D.~R.}\ \bibnamefont
  {Hamann}}\ and\ \bibinfo {author} {\bibfnamefont {D.}~\bibnamefont
  {Vanderbilt}},\ }\href {\doibase 10.1103/PhysRevB.79.045109} {\bibfield
  {journal} {\bibinfo  {journal} {Phys. Rev. B}\ }\textbf {\bibinfo {volume}
  {79}},\ \bibinfo {pages} {045109} (\bibinfo {year} {2009})}\BibitemShut
  {NoStop}%
\bibitem [{\citenamefont {Cappellini}\ \emph {et~al.}(2000)\citenamefont
  {Cappellini}, \citenamefont {Bouette-Russo}, \citenamefont {Amadon},
  \citenamefont {Noguera},\ and\ \citenamefont {Finocchi}}]{Cappellini00}%
  \BibitemOpen
  \bibfield  {author} {\bibinfo {author} {\bibfnamefont {G.}~\bibnamefont
  {Cappellini}}, \bibinfo {author} {\bibfnamefont {S.}~\bibnamefont
  {Bouette-Russo}}, \bibinfo {author} {\bibfnamefont {B.}~\bibnamefont
  {Amadon}}, \bibinfo {author} {\bibfnamefont {C.}~\bibnamefont {Noguera}}, \
  and\ \bibinfo {author} {\bibfnamefont {F.}~\bibnamefont {Finocchi}},\
  }\href@noop {} {\bibfield  {journal} {\bibinfo  {journal} {Journal of
  Physics: Condensed Matter}\ }\textbf {\bibinfo {volume} {12}},\ \bibinfo
  {pages} {3671} (\bibinfo {year} {2000})}\BibitemShut {NoStop}%
\bibitem [{\citenamefont {Kotani}\ \emph
  {et~al.}(2007{\natexlab{b}})\citenamefont {Kotani}, \citenamefont {van
  Schilfgaarde}, \citenamefont {Faleev},\ and\ \citenamefont
  {Chantis}}]{Takao07}%
  \BibitemOpen
  \bibfield  {author} {\bibinfo {author} {\bibfnamefont {T.}~\bibnamefont
  {Kotani}}, \bibinfo {author} {\bibfnamefont {M.}~\bibnamefont {van
  Schilfgaarde}}, \bibinfo {author} {\bibfnamefont {S.~V.}\ \bibnamefont
  {Faleev}}, \ and\ \bibinfo {author} {\bibfnamefont {A.}~\bibnamefont
  {Chantis}},\ }\href {http://stacks.iop.org/0953-8984/19/i=36/a=365236}
  {\bibfield  {journal} {\bibinfo  {journal} {Journal of Physics: Condensed
  Matter}\ }\textbf {\bibinfo {volume} {19}},\ \bibinfo {pages} {365236}
  (\bibinfo {year} {2007}{\natexlab{b}})}\BibitemShut {NoStop}%
\bibitem [{\citenamefont {Deguchi}\ \emph {et~al.}(2016)\citenamefont
  {Deguchi}, \citenamefont {Sato}, \citenamefont {Kino},\ and\ \citenamefont
  {Kotani}}]{Deguchi16}%
  \BibitemOpen
  \bibfield  {author} {\bibinfo {author} {\bibfnamefont {D.}~\bibnamefont
  {Deguchi}}, \bibinfo {author} {\bibfnamefont {K.}~\bibnamefont {Sato}},
  \bibinfo {author} {\bibfnamefont {H.}~\bibnamefont {Kino}}, \ and\ \bibinfo
  {author} {\bibfnamefont {T.}~\bibnamefont {Kotani}},\ }\href
  {http://stacks.iop.org/1347-4065/55/i=5/a=051201} {\bibfield  {journal}
  {\bibinfo  {journal} {Jap. J. Appl. Phys.}\ }\textbf {\bibinfo {volume}
  {55}},\ \bibinfo {pages} {051201} (\bibinfo {year} {2016})}\BibitemShut
  {NoStop}%
\bibitem [{\citenamefont {Ismail-Beigi}(2017)}]{Ismail-Beigi17}%
  \BibitemOpen
  \bibfield  {author} {\bibinfo {author} {\bibfnamefont {S.}~\bibnamefont
  {Ismail-Beigi}},\ }\href {http://stacks.iop.org/0953-8984/29/i=38/a=385501}
  {\bibfield  {journal} {\bibinfo  {journal} {Journal of Physics: Condensed
  Matter}\ }\textbf {\bibinfo {volume} {29}},\ \bibinfo {pages} {385501}
  (\bibinfo {year} {2017})}\BibitemShut {NoStop}%
\bibitem [{\citenamefont {Chen}\ and\ \citenamefont
  {Pasquarello}(2015)}]{PasquarelloChen15}%
  \BibitemOpen
  \bibfield  {author} {\bibinfo {author} {\bibfnamefont {W.}~\bibnamefont
  {Chen}}\ and\ \bibinfo {author} {\bibfnamefont {A.}~\bibnamefont
  {Pasquarello}},\ }\href {\doibase 10.1103/PhysRevB.92.041115} {\bibfield
  {journal} {\bibinfo  {journal} {Phys. Rev. B}\ }\textbf {\bibinfo {volume}
  {92}},\ \bibinfo {pages} {041115} (\bibinfo {year} {2015})}\BibitemShut
  {NoStop}%
\bibitem [{\citenamefont {Kutepov}(2017)}]{Kutepov17}%
  \BibitemOpen
  \bibfield  {author} {\bibinfo {author} {\bibfnamefont {A.~L.}\ \bibnamefont
  {Kutepov}},\ }\href {\doibase 10.1103/PhysRevB.95.195120} {\bibfield
  {journal} {\bibinfo  {journal} {Phys. Rev. B}\ }\textbf {\bibinfo {volume}
  {95}},\ \bibinfo {pages} {195120} (\bibinfo {year} {2017})}\BibitemShut
  {NoStop}%
\bibitem [{\citenamefont {Shishkin}\ \emph {et~al.}(2007)\citenamefont
  {Shishkin}, \citenamefont {Marsman},\ and\ \citenamefont
  {Kresse}}]{Shishkin07}%
  \BibitemOpen
  \bibfield  {author} {\bibinfo {author} {\bibfnamefont {M.}~\bibnamefont
  {Shishkin}}, \bibinfo {author} {\bibfnamefont {M.}~\bibnamefont {Marsman}}, \
  and\ \bibinfo {author} {\bibfnamefont {G.}~\bibnamefont {Kresse}},\ }\href
  {\doibase 10.1103/PhysRevLett.99.246403} {\bibfield  {journal} {\bibinfo
  {journal} {Phys. Rev. Lett.}\ }\textbf {\bibinfo {volume} {99}},\ \bibinfo
  {pages} {246403} (\bibinfo {year} {2007})}\BibitemShut {NoStop}%
\bibitem [{\citenamefont {Chantis}\ \emph {et~al.}(2006)\citenamefont
  {Chantis}, \citenamefont {van Schilfgaarde},\ and\ \citenamefont
  {Kotani}}]{Chantis06}%
  \BibitemOpen
  \bibfield  {author} {\bibinfo {author} {\bibfnamefont {A.~N.}\ \bibnamefont
  {Chantis}}, \bibinfo {author} {\bibfnamefont {M.}~\bibnamefont {van
  Schilfgaarde}}, \ and\ \bibinfo {author} {\bibfnamefont {T.}~\bibnamefont
  {Kotani}},\ }\href {\doibase 10.1103/PhysRevLett.96.086405} {\bibfield
  {journal} {\bibinfo  {journal} {Phys. Rev. Lett.}\ }\textbf {\bibinfo
  {volume} {96}},\ \bibinfo {pages} {086405} (\bibinfo {year}
  {2006})}\BibitemShut {NoStop}%
\bibitem [{\citenamefont {Chantis}\ \emph {et~al.}(2008)\citenamefont
  {Chantis}, \citenamefont {Cardona}, \citenamefont {Christensen},
  \citenamefont {Smith}, \citenamefont {van Schilfgaarde}, \citenamefont
  {Kotani}, \citenamefont {Svane},\ and\ \citenamefont {Albers}}]{Chantis08}%
  \BibitemOpen
  \bibfield  {author} {\bibinfo {author} {\bibfnamefont {A.~N.}\ \bibnamefont
  {Chantis}}, \bibinfo {author} {\bibfnamefont {M.}~\bibnamefont {Cardona}},
  \bibinfo {author} {\bibfnamefont {N.~E.}\ \bibnamefont {Christensen}},
  \bibinfo {author} {\bibfnamefont {D.~L.}\ \bibnamefont {Smith}}, \bibinfo
  {author} {\bibfnamefont {M.}~\bibnamefont {van Schilfgaarde}}, \bibinfo
  {author} {\bibfnamefont {T.}~\bibnamefont {Kotani}}, \bibinfo {author}
  {\bibfnamefont {A.}~\bibnamefont {Svane}}, \ and\ \bibinfo {author}
  {\bibfnamefont {R.~C.}\ \bibnamefont {Albers}},\ }\href {\doibase
  10.1103/PhysRevB.78.075208} {\bibfield  {journal} {\bibinfo  {journal} {Phys.
  Rev. B}\ }\textbf {\bibinfo {volume} {78}},\ \bibinfo {pages} {075208}
  (\bibinfo {year} {2008})}\BibitemShut {NoStop}%
\bibitem [{\citenamefont {Rigamonti}\ \emph {et~al.}(2015)\citenamefont
  {Rigamonti}, \citenamefont {Botti}, \citenamefont {Veniard}, \citenamefont
  {Draxl}, \citenamefont {Reining},\ and\ \citenamefont
  {Sottile}}]{Rigamonti15}%
  \BibitemOpen
  \bibfield  {author} {\bibinfo {author} {\bibfnamefont {S.}~\bibnamefont
  {Rigamonti}}, \bibinfo {author} {\bibfnamefont {S.}~\bibnamefont {Botti}},
  \bibinfo {author} {\bibfnamefont {V.}~\bibnamefont {Veniard}}, \bibinfo
  {author} {\bibfnamefont {C.}~\bibnamefont {Draxl}}, \bibinfo {author}
  {\bibfnamefont {L.}~\bibnamefont {Reining}}, \ and\ \bibinfo {author}
  {\bibfnamefont {F.}~\bibnamefont {Sottile}},\ }\href {\doibase
  10.1103/PhysRevLett.114.146402} {\bibfield  {journal} {\bibinfo  {journal}
  {Phys. Rev. Lett.}\ }\textbf {\bibinfo {volume} {114}},\ \bibinfo {pages}
  {146402} (\bibinfo {year} {2015})}\BibitemShut {NoStop}%
\bibitem [{\citenamefont {Kutepov}(2016)}]{Kutepov16}%
  \BibitemOpen
  \bibfield  {author} {\bibinfo {author} {\bibfnamefont {A.~L.}\ \bibnamefont
  {Kutepov}},\ }\href {\doibase 10.1103/PhysRevB.94.155101} {\bibfield
  {journal} {\bibinfo  {journal} {Phys. Rev. B}\ }\textbf {\bibinfo {volume}
  {94}},\ \bibinfo {pages} {155101} (\bibinfo {year} {2016})}\BibitemShut
  {NoStop}%
\bibitem [{\citenamefont {Hedin}(1965)}]{Hedin65}%
  \BibitemOpen
  \bibfield  {author} {\bibinfo {author} {\bibfnamefont {L.}~\bibnamefont
  {Hedin}},\ }\href {\doibase 10.1103/PhysRev.139.A796} {\bibfield  {journal}
  {\bibinfo  {journal} {Phys. Rev.}\ }\textbf {\bibinfo {volume} {139}},\
  \bibinfo {pages} {A796} (\bibinfo {year} {1965})}\BibitemShut {NoStop}%
\bibitem [{\citenamefont {Hedin}\ and\ \citenamefont
  {Lundqvist}(1969)}]{Hedin69}%
  \BibitemOpen
  \bibfield  {author} {\bibinfo {author} {\bibfnamefont {L.}~\bibnamefont
  {Hedin}}\ and\ \bibinfo {author} {\bibfnamefont {S.}~\bibnamefont
  {Lundqvist}},\ }in\ \href@noop {} {\emph {\bibinfo {booktitle} {Solid State
  Physics, Advanced in Research and Applications}}},\ Vol.~\bibinfo {volume}
  {23},\ \bibinfo {editor} {edited by\ \bibinfo {editor} {\bibfnamefont
  {F.}~\bibnamefont {Seitz}}, \bibinfo {editor} {\bibfnamefont
  {D.}~\bibnamefont {Turnbull}}, \ and\ \bibinfo {editor} {\bibfnamefont
  {H.}~\bibnamefont {Ehrenreich}}}\ (\bibinfo  {publisher} {Academic Press},\
  \bibinfo {address} {New York},\ \bibinfo {year} {1969})\ pp.\ \bibinfo
  {pages} {1--181}\BibitemShut {NoStop}%
\bibitem [{\citenamefont {Lambrecht}\ \emph {et~al.}(2017)\citenamefont
  {Lambrecht}, \citenamefont {Bhandari},\ and\ \citenamefont {van
  Schilfgaarde}}]{Lambrecht17}%
  \BibitemOpen
  \bibfield  {author} {\bibinfo {author} {\bibfnamefont {W.~R.~L.}\
  \bibnamefont {Lambrecht}}, \bibinfo {author} {\bibfnamefont {C.}~\bibnamefont
  {Bhandari}}, \ and\ \bibinfo {author} {\bibfnamefont {M.}~\bibnamefont {van
  Schilfgaarde}},\ }\href {\doibase 10.1103/PhysRevMaterials.1.043802}
  {\bibfield  {journal} {\bibinfo  {journal} {Phys. Rev. Materials}\ }\textbf
  {\bibinfo {volume} {1}},\ \bibinfo {pages} {043802} (\bibinfo {year}
  {2017})}\BibitemShut {NoStop}%
\bibitem [{\citenamefont {Botti}\ and\ \citenamefont
  {Marques}(2013)}]{Botti13}%
  \BibitemOpen
  \bibfield  {author} {\bibinfo {author} {\bibfnamefont {S.}~\bibnamefont
  {Botti}}\ and\ \bibinfo {author} {\bibfnamefont {M.~A.~L.}\ \bibnamefont
  {Marques}},\ }\href {\doibase 10.1103/PhysRevLett.110.226404} {\bibfield
  {journal} {\bibinfo  {journal} {Phys. Rev. Lett.}\ }\textbf {\bibinfo
  {volume} {110}},\ \bibinfo {pages} {226404} (\bibinfo {year}
  {2013})}\BibitemShut {NoStop}%
\bibitem [{\citenamefont {Friedrich}\ \emph {et~al.}(2010)\citenamefont
  {Friedrich}, \citenamefont {Bl\"ugel},\ and\ \citenamefont
  {Schindlmayr}}]{Friedrich10}%
  \BibitemOpen
  \bibfield  {author} {\bibinfo {author} {\bibfnamefont {C.}~\bibnamefont
  {Friedrich}}, \bibinfo {author} {\bibfnamefont {S.}~\bibnamefont {Bl\"ugel}},
  \ and\ \bibinfo {author} {\bibfnamefont {A.}~\bibnamefont {Schindlmayr}},\
  }\href {\doibase 10.1103/PhysRevB.81.125102} {\bibfield  {journal} {\bibinfo
  {journal} {Phys. Rev. B}\ }\textbf {\bibinfo {volume} {81}},\ \bibinfo
  {pages} {125102} (\bibinfo {year} {2010})}\BibitemShut {NoStop}%
\bibitem [{\citenamefont {Kotani}(2014)}]{Kotanijpsj}%
  \BibitemOpen
  \bibfield  {author} {\bibinfo {author} {\bibfnamefont {T.}~\bibnamefont
  {Kotani}},\ }\href {\doibase 10.7566/JPSJ.83.094711} {\bibfield  {journal}
  {\bibinfo  {journal} {Journal of the Physical Society of Japan}\ }\textbf
  {\bibinfo {volume} {83}},\ \bibinfo {pages} {094711} (\bibinfo {year}
  {2014})},\ \Eprint
  {http://arxiv.org/abs/http://dx.doi.org/10.7566/JPSJ.83.094711}
  {http://dx.doi.org/10.7566/JPSJ.83.094711} \BibitemShut {NoStop}%
\bibitem [{que()}]{questaal}%
  \BibitemOpen
  \href@noop {} {}\bibinfo {howpublished} {\url{https://www.questaal.org/}, Our
  $GW$ implementation was adapted from the original ecalj package now at
  \url{https://github.com/tkotani/ecalj/}.}\BibitemShut {Stop}%
\bibitem [{\citenamefont {Bott}\ \emph {et~al.}(1998)\citenamefont {Bott},
  \citenamefont {Methfessel}, \citenamefont {Krabs},\ and\ \citenamefont
  {Schmidt}}]{Bott98}%
  \BibitemOpen
  \bibfield  {author} {\bibinfo {author} {\bibfnamefont {E.}~\bibnamefont
  {Bott}}, \bibinfo {author} {\bibfnamefont {M.}~\bibnamefont {Methfessel}},
  \bibinfo {author} {\bibfnamefont {W.}~\bibnamefont {Krabs}}, \ and\ \bibinfo
  {author} {\bibfnamefont {P.~C.}\ \bibnamefont {Schmidt}},\ }\href {\doibase
  10.1063/1.532437} {\bibfield  {journal} {\bibinfo  {journal} {Journal of
  Mathematical Physics}\ }\textbf {\bibinfo {volume} {39}},\ \bibinfo {pages}
  {3393} (\bibinfo {year} {1998})}\BibitemShut {NoStop}%
\bibitem [{\citenamefont {Cabaret}\ \emph {et~al.}(2007)\citenamefont
  {Cabaret}, \citenamefont {Couzinet}, \citenamefont {Flank}, \citenamefont
  {Itié}, \citenamefont {Lagarde},\ and\ \citenamefont {Polian}}]{Cabaret07}%
  \BibitemOpen
  \bibfield  {author} {\bibinfo {author} {\bibfnamefont {D.}~\bibnamefont
  {Cabaret}}, \bibinfo {author} {\bibfnamefont {B.}~\bibnamefont {Couzinet}},
  \bibinfo {author} {\bibfnamefont {A.}~\bibnamefont {Flank}}, \bibinfo
  {author} {\bibfnamefont {J.}~\bibnamefont {Itié}}, \bibinfo {author}
  {\bibfnamefont {P.}~\bibnamefont {Lagarde}}, \ and\ \bibinfo {author}
  {\bibfnamefont {A.}~\bibnamefont {Polian}},\ }\href {\doibase
  http://dx.doi.org/10.1063/1.2644447} {\bibfield  {journal} {\bibinfo
  {journal} {AIP Conference Proceedings}\ }\textbf {\bibinfo {volume} {882}},\
  \bibinfo {pages} {120} (\bibinfo {year} {2007})}\BibitemShut {NoStop}%
\bibitem [{\citenamefont {Bhandari}\ and\ \citenamefont
  {Lambrecht}(2017)}]{Bhandaristostruc}%
  \BibitemOpen
  \bibfield  {author} {\bibinfo {author} {\bibfnamefont {C.}~\bibnamefont
  {Bhandari}}\ and\ \bibinfo {author} {\bibfnamefont {W.~R.~L.}\ \bibnamefont
  {Lambrecht}},\ }\href@noop {} {\enquote {\bibinfo {title} {{Instability of
  the layered orthorhombic post-perovskite phase of SrTiO$_3$}},}\ } (\bibinfo
  {year} {2017}),\ \bibinfo {note} {unpublished}\BibitemShut {NoStop}%
\bibitem [{\citenamefont {Abramov}\ \emph {et~al.}(1995)\citenamefont
  {Abramov}, \citenamefont {Tsirelson}, \citenamefont {Zavodnik}, \citenamefont
  {Ivanov},\ and\ \citenamefont {D.}}]{Abramov95}%
  \BibitemOpen
  \bibfield  {author} {\bibinfo {author} {\bibfnamefont {Y.~A.}\ \bibnamefont
  {Abramov}}, \bibinfo {author} {\bibfnamefont {V.~G.}\ \bibnamefont
  {Tsirelson}}, \bibinfo {author} {\bibfnamefont {V.~E.}\ \bibnamefont
  {Zavodnik}}, \bibinfo {author} {\bibfnamefont {S.~A.}\ \bibnamefont
  {Ivanov}}, \ and\ \bibinfo {author} {\bibfnamefont {B.~I.}\ \bibnamefont
  {D.}},\ }\href {\doibase 10.1107/S0108768195003752} {\bibfield  {journal}
  {\bibinfo  {journal} {Acta Crystallographica Section B}\ }\textbf {\bibinfo
  {volume} {51}},\ \bibinfo {pages} {942} (\bibinfo {year} {1995})}\BibitemShut
  {NoStop}%
\bibitem [{\citenamefont {Jauch}\ and\ \citenamefont {Palmer}(1999)}]{Jauch99}%
  \BibitemOpen
  \bibfield  {author} {\bibinfo {author} {\bibfnamefont {W.}~\bibnamefont
  {Jauch}}\ and\ \bibinfo {author} {\bibfnamefont {A.}~\bibnamefont {Palmer}},\
  }\href {\doibase 10.1103/PhysRevB.60.2961} {\bibfield  {journal} {\bibinfo
  {journal} {Phys. Rev. B}\ }\textbf {\bibinfo {volume} {60}},\ \bibinfo
  {pages} {2961} (\bibinfo {year} {1999})}\BibitemShut {NoStop}%
\bibitem [{\citenamefont {Cai}\ \emph {et~al.}(2004)\citenamefont {Cai},
  \citenamefont {Yin},\ and\ \citenamefont {Zhang}}]{Meng04}%
  \BibitemOpen
  \bibfield  {author} {\bibinfo {author} {\bibfnamefont {M.-Q.}\ \bibnamefont
  {Cai}}, \bibinfo {author} {\bibfnamefont {Z.}~\bibnamefont {Yin}}, \ and\
  \bibinfo {author} {\bibfnamefont {M.-S.}\ \bibnamefont {Zhang}},\ }\href
  {\doibase http://dx.doi.org/10.1016/j.cplett.2004.02.095} {\bibfield
  {journal} {\bibinfo  {journal} {Chemical Physics Letters}\ }\textbf {\bibinfo
  {volume} {388}},\ \bibinfo {pages} {223 } (\bibinfo {year}
  {2004})}\BibitemShut {NoStop}%
\bibitem [{\citenamefont {Gupta}\ \emph {et~al.}(2004)\citenamefont {Gupta},
  \citenamefont {Nautiyal},\ and\ \citenamefont {Auluck}}]{Gupta04}%
  \BibitemOpen
  \bibfield  {author} {\bibinfo {author} {\bibfnamefont {G.}~\bibnamefont
  {Gupta}}, \bibinfo {author} {\bibfnamefont {T.}~\bibnamefont {Nautiyal}}, \
  and\ \bibinfo {author} {\bibfnamefont {S.}~\bibnamefont {Auluck}},\ }\href
  {\doibase 10.1103/PhysRevB.69.052101} {\bibfield  {journal} {\bibinfo
  {journal} {Phys. Rev. B}\ }\textbf {\bibinfo {volume} {69}},\ \bibinfo
  {pages} {052101} (\bibinfo {year} {2004})}\BibitemShut {NoStop}%
\bibitem [{\citenamefont {van Benthem}\ \emph {et~al.}(2001)\citenamefont {van
  Benthem}, \citenamefont {Elsässer},\ and\ \citenamefont {French}}]{van01}%
  \BibitemOpen
  \bibfield  {author} {\bibinfo {author} {\bibfnamefont {K.}~\bibnamefont {van
  Benthem}}, \bibinfo {author} {\bibfnamefont {C.}~\bibnamefont {Elsässer}}, \
  and\ \bibinfo {author} {\bibfnamefont {R.~H.}\ \bibnamefont {French}},\
  }\href {\doibase http://dx.doi.org/10.1063/1.1415766} {\bibfield  {journal}
  {\bibinfo  {journal} {J. Appl. Phys.}\ }\textbf {\bibinfo {volume} {90}},\
  \bibinfo {pages} {6156} (\bibinfo {year} {2001})}\BibitemShut {NoStop}%
\bibitem [{\citenamefont {Bäuerle}\ \emph {et~al.}(1978)\citenamefont
  {Bäuerle}, \citenamefont {Braun}, \citenamefont {Saile}, \citenamefont
  {Sprüssel},\ and\ \citenamefont {Koch}}]{Bauerle78}%
  \BibitemOpen
  \bibfield  {author} {\bibinfo {author} {\bibfnamefont {D.}~\bibnamefont
  {Bäuerle}}, \bibinfo {author} {\bibfnamefont {W.}~\bibnamefont {Braun}},
  \bibinfo {author} {\bibfnamefont {V.}~\bibnamefont {Saile}}, \bibinfo
  {author} {\bibfnamefont {G.}~\bibnamefont {Sprüssel}}, \ and\ \bibinfo
  {author} {\bibfnamefont {E.}~\bibnamefont {Koch}},\ }\href {\doibase
  10.1007/BF01321179} {\bibfield  {journal} {\bibinfo  {journal} {Zeitschrift
  für Physik B Condensed Matter}\ }\textbf {\bibinfo {volume} {29}},\ \bibinfo
  {pages} {179} (\bibinfo {year} {1978})}\BibitemShut {NoStop}%
\bibitem [{\citenamefont {Cardona}(1965)}]{Cardona65}%
  \BibitemOpen
  \bibfield  {author} {\bibinfo {author} {\bibfnamefont {M.}~\bibnamefont
  {Cardona}},\ }\href {\doibase 10.1103/PhysRev.140.A651} {\bibfield  {journal}
  {\bibinfo  {journal} {Phys. Rev.}\ }\textbf {\bibinfo {volume} {140}},\
  \bibinfo {pages} {A651} (\bibinfo {year} {1965})}\BibitemShut {NoStop}%
\bibitem [{\citenamefont {Nery}\ and\ \citenamefont
  {Allen}(2016)}]{NeryAllen16}%
  \BibitemOpen
  \bibfield  {author} {\bibinfo {author} {\bibfnamefont {J.~P.}\ \bibnamefont
  {Nery}}\ and\ \bibinfo {author} {\bibfnamefont {P.~B.}\ \bibnamefont
  {Allen}},\ }\href {\doibase 10.1103/PhysRevB.94.115135} {\bibfield  {journal}
  {\bibinfo  {journal} {Phys. Rev. B}\ }\textbf {\bibinfo {volume} {94}},\
  \bibinfo {pages} {115135} (\bibinfo {year} {2016})}\BibitemShut {NoStop}%
\bibitem [{\citenamefont {Janotti}\ \emph {et~al.}(2014)\citenamefont
  {Janotti}, \citenamefont {Varley}, \citenamefont {Choi},\ and\ \citenamefont
  {Van~de Walle}}]{Janotti14}%
  \BibitemOpen
  \bibfield  {author} {\bibinfo {author} {\bibfnamefont {A.}~\bibnamefont
  {Janotti}}, \bibinfo {author} {\bibfnamefont {J.~B.}\ \bibnamefont {Varley}},
  \bibinfo {author} {\bibfnamefont {M.}~\bibnamefont {Choi}}, \ and\ \bibinfo
  {author} {\bibfnamefont {C.~G.}\ \bibnamefont {Van~de Walle}},\ }\href
  {\doibase 10.1103/PhysRevB.90.085202} {\bibfield  {journal} {\bibinfo
  {journal} {Phys. Rev. B}\ }\textbf {\bibinfo {volume} {90}},\ \bibinfo
  {pages} {085202} (\bibinfo {year} {2014})}\BibitemShut {NoStop}%
\bibitem [{\citenamefont {Erhart}\ \emph {et~al.}(2014)\citenamefont {Erhart},
  \citenamefont {Klein}, \citenamefont {\AA{}berg},\ and\ \citenamefont
  {Sadigh}}]{Erhart14}%
  \BibitemOpen
  \bibfield  {author} {\bibinfo {author} {\bibfnamefont {P.}~\bibnamefont
  {Erhart}}, \bibinfo {author} {\bibfnamefont {A.}~\bibnamefont {Klein}},
  \bibinfo {author} {\bibfnamefont {D.}~\bibnamefont {\AA{}berg}}, \ and\
  \bibinfo {author} {\bibfnamefont {B.}~\bibnamefont {Sadigh}},\ }\href
  {\doibase 10.1103/PhysRevB.90.035204} {\bibfield  {journal} {\bibinfo
  {journal} {Phys. Rev. B}\ }\textbf {\bibinfo {volume} {90}},\ \bibinfo
  {pages} {035204} (\bibinfo {year} {2014})}\BibitemShut {NoStop}%
\bibitem [{\citenamefont {Heifets}\ \emph {et~al.}(2006)\citenamefont
  {Heifets}, \citenamefont {Kotomin},\ and\ \citenamefont
  {Trepakov}}]{Heifets06}%
  \BibitemOpen
  \bibfield  {author} {\bibinfo {author} {\bibfnamefont {E.}~\bibnamefont
  {Heifets}}, \bibinfo {author} {\bibfnamefont {E.}~\bibnamefont {Kotomin}}, \
  and\ \bibinfo {author} {\bibfnamefont {V.~A.}\ \bibnamefont {Trepakov}},\
  }\href {http://stacks.iop.org/0953-8984/18/i=20/a=009} {\bibfield  {journal}
  {\bibinfo  {journal} {Journal of Physics: Condensed Matter}\ }\textbf
  {\bibinfo {volume} {18}},\ \bibinfo {pages} {4845} (\bibinfo {year}
  {2006})}\BibitemShut {NoStop}%
\end{thebibliography}%

\end{document}